\documentclass[a4paper]{article}
\usepackage[a4paper,top=3cm,bottom=2cm,left=3cm,right=3cm,marginparwidth=1.75cm]{geometry}
\usepackage{amsfonts}
\usepackage{bm}
\usepackage{extarrows}
\usepackage{amssymb}
\usepackage{color}
\usepackage[all]{xy}
\usepackage{graphicx}
\usepackage{subfig}
\usepackage{braket}
\usepackage{amsmath}
\usepackage{appendix}
\usepackage{graphicx}
\usepackage[colorinlistoftodos]{todonotes}
\usepackage{amsthm}
\usepackage{mathrsfs}
\usepackage{amssymb}
\usepackage{accents}
\usepackage{extarrows}
\usepackage{makecell}
\usepackage{authblk}
\usepackage{url}
\usepackage{hyperref}
\usepackage{cite}
\usepackage{eufrak}
\hypersetup{colorlinks,
            citecolor=black,
            linkcolor=black,
            urlcolor=green,
            pdftex}

\newtheorem{theorem}{Theorem}

\usepackage[english]{babel}
\usepackage[utf8x]{inputenc}
\usepackage[T1]{fontenc}

\usepackage[normalem]{ulem}

\allowdisplaybreaks[4]

\usepackage{enumitem}

\title{Entanglement entropy of coherent intertwiner in loop quantum gravity}
\author[1]{Gaoping Long \footnote{201731140005@mail.bnu.edu.cn}}
\author[2]{Qian Chen \footnote{chenqian.phys@gmail.com}\thanks{corresponding author}}
\author[3]{Jinsong Yang\footnote{jsyang@gzu.edu.cn}}
\affil[1]{College of Physics $\&$ Optoelectronic Engineering, Jinan University, Guangzhou, 510632, Guangdong, China}
\affil[2]{Univ Lyon, Inria, ENS Lyon, UCBL, LIP, F-69342, Lyon Cedex 07, France}
\affil[3]{School of Physics, Guizhou University, Guiyang 550025, China}

\date{}

\begin{document}

\maketitle

\begin{abstract}
In this paper, we carry out the entanglement calculations on the coherent intertwiners.  We first consider  the entanglement  introduced by the group-averaging of the tensor-product type intertwiner on a four-valents vertex. The result shows that the entanglement is determined by the probability distribution of recoupling spin, and this probability distribution is a well-behaved peak for the highest (and lowest) weight states.  Further, we calculated explicitly the entanglement on gauge-invariant coherent intertwiner with four legs. Our numerical results show that the shape of the semiclassical polyhedron described by the coherent intertwiner can be related to the entanglement; In other words, the entanglement is controlled by the face-angle of the  semiclassical polyhedron. Finally, we extend our analytical calculation to  the coherent intertwiners with arbitrary number of legs. 

\end{abstract}
\section{Introduction}
Though loop quantum gravity (LQG) provides a background-independent and non-perturbative quantum theory of General Relativity (GR) \cite{30years,Ashtekar:2004eh,thiemann2008modern,rovelli_vidotto_2014,rovelli2007quantum,Han2005FUNDAMENTAL},  the reconstruction of classical geometry in LQG is an unsettled crucial concern. Specifically, LQG defines the quantum states of spatial intrinsic and extrinsic geometry as spin-networks. The quantum  geometry carried by spin-networks can  be understood as the quantization of discrete twisted geometries \cite{Rovelli:2010km,Freidel:2010bw,PhysRevD.82.084040,PhysRevD.103.086016,Long:2023ivt}, which is a generalization of the Regge geometry \cite{regge}. Then, the evolution of spin-networks is governed by the Hamiltonian operator in canonical formulation \cite{Alesci:2015wla,Assanioussi:2015gka,Zhang:2021qul,Yang:2015zda}, or described by  path-integral formulation \cite{Perez:2012wv,Han:2021kll,Han:2019vpw,Han:2020chr,Long:2021izw}. This description of quantum gravity provides a discrete picture of quantum geometry at Plank scale,  which indicates that the paradigm of classical smooth space-time geometry is just a semi-classical and large scale approximation. More explicitly, the information of quantum geometry encoded in spin-network states can be extracted by the geometric operators in LQG \cite{Ashtekar:1996eg,Ashtekar:1997fb,Ma:2010fy,Bianchi:2008es,long2020operators}; Correspondingly, the semi-classical geometries can be given by the expectation values of the geometric operators based on the coherent states, which are constructed by specific superposition of spin-networks \cite{Thomas2001Gauge,2001Gauge,Bianchi:2009ky,Calcinari_2020}. However, the resulting semi-classical geometry takes the discrete formulation of twisted geometry, and they still need to be coarse-grained in order to derive the structure of smooth classical geometry \cite{Livine:2006xk}. This issue is a quite hard problem in LQG, so that a dynamical description of smooth geometry from the evolution of spin-networks is still an open question.

Nevertheless, there is another perspective which could provides us a guidance to recognize the coarse-grain.  This idea is to study quantum statistical properties (e.g. the entanglement) on spin-networks, and establish the relation between them and the classical geometries (e.g. entropy-area law). In fact, the notion of quantum information is essential to a background independent quantum gravity: the relational perspective plays a very important role in the interpretation of the theory, which could be fully carried by the relations between local Hilbert spaces, namely, the structure of quantum entanglement.
The studies following this idea have been carried out in many works \cite{Donnelly:2008vx,Donnelly:2011hn,Donnelly:2014gva,Bianchi:2015fra,Delcamp:2016eya,Livine:2017fgq,Baytas:2018wjd,Chen:2022rty}. It has been shown that the (physical) entanglement carried by the spin-networks comes from the superpositions of spins and intertwiners. In fact, the physical states in LQG can be established by kinds of superpositions of spin-network basis. Thus, it is a question that which kind of  superpositions of spin-network could give a realistic relation between the quantum statistics and the geometries.
As mentioned above, by requiring the semi-classicality of the quantum geometry,  the coherent states of spin-networks provides some superpositions of the spin-networks naturally \cite{Thomas2001Gauge,2001Gauge,Bianchi:2009ky,Calcinari_2020,PhysRevD.104.046014,1994The,Long:2021lmd,Long:2022cex}. It is worth to explore the statistics based on such kind of superpositions of spin-networks, to relate them to the geometries at the semi-classical level.
In fact, a gauge-variant coherent state of spin-network on a closed oriented graph is given by the product of the heat-kernel coherent state of $SU(2)$ on each edge. Such state carries no intertwiner entanglement \cite{Livine:2017fgq}. Gauge averaging over  the product state generates the coherent intertwiners at vertices, which introduce superposition and correlation onto the state thus the entanglement can access into the play. In this paper, we will focus on the mechanism of group averaging to the entanglement, specifically, of coherent intertwiners, which could be a preparation for the further related studies on the coherent state entanglement.


This paper is organized as follows. After the basic structure of spin-network state and intertwiner in LQG  being introduced in section \ref{sec2},   the entanglements on kinds of spin-network state and  intertwiners are calculated.  Specifically, in section \ref{sec3.1}, we  generalize the relation between boundary entanglement and intertwiner entanglement  to the case that the internal edge  carries spin-superposition; Then,  we calculate the entanglement introduced by the group-averaging of the tensor-product type intertwiner with four legs by numerical method in section \ref{sec3.2}; Further, in section \ref{sec3.3.1}, this calculation is extended to the gauge-invariant coherent intertwiner with four legs; Moreover, in section \ref{sec3.3.2}, we also carry out some key analytical calculations for the entanglement on the coherent intertwiners with arbitrary number of legs. Finally, we finish with conclusion and outlook for further research in section \ref{sec4} .




\section{Spin-network state and intertwiner in loop quantum gravity}\label{sec2}
 The Hilbert space $\mathcal{H}_{\Gamma}$ for quantum geometry on  a closed oriented graph $\Gamma$ embedded in a 3-dimensional manifold is composed by the square integrable functions on $SU(2)$ associated to each edge $e\in\Gamma$, which are invariant under the $SU(2)$ action at every vertex $v\in\Gamma$.
 Specifically, a square integrable functions on $\Gamma$ takes the formulation
\begin{equation}
\Psi_\Gamma=\Psi_\Gamma(\{h_{e}\}_{e\in\Gamma}).
\end{equation}
The $SU(2)$ gauge invariance at the vertex of $\Psi_\Gamma$ reads
\begin{equation}
\Psi_\Gamma(\{h_{e}\}_{e\in\Gamma})=\Psi_\Gamma(\{g_{s(e)}h_{e}g^{-1}_{t(e)}\}_{e\in\Gamma}),
\end{equation}
where $g_v$ is given at each vertex $v\in\Gamma$ respectively, $s(e)$ represents the source vertex of $e$ and
$t(e)$ the target vertex of $e$. The spin-network states provide a basis of space $\mathcal{H}_\Gamma$. Specifically, a spin-network state on $\Gamma$ is given by labeling a spin $j_e\in\frac{\mathbb{N}}{2}$ on each edge $e\in\Gamma$ and an intertwiner $\mathcal{I}_v$ on each vertex $v\in\Gamma$, which reads \cite{Ashtekar:2004eh} 
\begin{equation}
\Psi_{\Gamma,\{j_e,\mathcal{I}_v\}}=\text{tr}\left(\bigotimes_{e\in\Gamma}\pi_{j_e}(h_e)\bigotimes_{v\in\Gamma} \mathcal{I}_v \right)
\end{equation}
where $\pi_{j_e}(h_e)$ is the representation matrix of $h_e\in SU(2)$ in the representation space $V^{j_e}$ of $SU(2)$ labelled by spin $j_e$, and  $\mathcal{I}_v\in\bigotimes_{e|t(e)=v} V^{j_e}
\otimes
\bigotimes_{e|s(e)=v} \bar{V}^{j_e}$. Especially, the spin-network state $\Psi_{\Gamma,\{j_e,\mathcal{I}_v\}}$ is gauge invariant if and only if each $v\in\Gamma$ is labelled by a gauge invariant intertwiner.  Another basis of space $\mathcal{H}_\Gamma$ is given  by the coherent state of spin-networks.  The coherent state of spin-network is the superposition of spin-networks, which reads \cite{Thomas2001Gauge}
\begin{equation}\label{heat}
\Psi_{\Gamma,{G}}^{{t}}({h})=\prod_{e\in \Gamma}\Psi^{t}_{G_e}(h_e)
\end{equation}
with
\begin{equation}
\Psi^{t}_{G_e}(h_e):=\sum_{j_e\in\frac{\mathbb{N}_{\scriptscriptstyle +}}{2}}(2j_e+1)e^{-tj_e(j_e+1)/2}\chi_{j_e}( h_e G_e^{-1}),
\end{equation}
where ${G}=\{{G}_e\}_{e\in \Gamma}$, ${h}=\{{h}_e\}_{e\in \Gamma}$, $\chi_{j}$ is the $SU(2)$ character with spin $j$ and $t\propto\kappa\hbar$ is a semi-classicality parameter. As a function of the holonomies $h_e$, the coherent state is labelled by  $G_e$ , with $G_e\in T^\ast SU(2)\cong SL(2,\mathbb{C})$ being the complex coordinates of the discrete holonomy-flux phase space of LQG.   The gauge invariant coherent state of spin-network is labelled by gauge equivalent class of $G_e\sim G^{g}_e:=g^{-1}_{s(e)}G_eg_{t(e)}$ for all $e\in \Gamma$, where $g=\{g_v\in SU(2)|v\in \Gamma\}$ . Equivalently, the gauge invariant coherent state of spin-network is also labelled by the gauge invariant intertwiners at each $v\in\Gamma$. Let us give an explicit introduction of gauge invariant intertwiner as follows.

The gauge invariant intertwiner $\mathcal{I}_v$ at vertex $v$ is a $SU(2)$-invariant state  in the tensor product space of all the spins associated to the edges linked to $v$,
\begin{equation}
\mathcal{I}_v\in\mathcal{H}_v^{\{j_e\}}:= \text{Inv}_{SU(2)}\left[\bigotimes_{e|t(e)=v} V^{j_e}
\otimes
\bigotimes_{e|s(e)=v} \bar{V}^{j_e}\right],
\end{equation}
where $\bar{V}^{j}$ is the dual space of $V^{j}$. The space $V^j$ has dimension $d_j=(2j+1)$ and the orthonormal basis $\{|j,m\rangle|-j\leq m\leq j\}$, which diagonalize the $su(2)$ Casimir $\vec{J}^2:=J^iJ_i, i=1,2,3$ and the generator $J_3$ as
\begin{equation}
\vec{J}^2|j,m\rangle=j(j+1)|j,m\rangle,\quad J_3|j,m\rangle=m|j,m\rangle.
\end{equation}
An orthonormal basis of the intertwiner space $\mathcal{H}_v^{\{j_e\}}$ is established by the recoupling scheme, which reads
\begin{equation}
\{\mathcal{I}_{v,\{j_\imath\}}^{\{j_e\}}=\mathcal{I}_{v,\{j_{\imath_1},j_{\imath_2},...,j_{\imath_{N_v-3}}\}}^{\{j_e\}}\in \mathcal{H}_v^{\{j_e\}}\},
\end{equation}
where $N_v$ is the number of the edges which link to $v$, and ${\{j_{\imath_1},j_{\imath_2},...,j_{\imath_{N_v-3}}\}}$ labeling the internal edges in the recoupling scheme, which satisfies
\begin{eqnarray}
&&|j_{e_1}-j_{e_2}|\leq j_{\imath_1}\leq j_{e_1}+j_{e_2},\quad
|j_{\imath_1}-j_{e_3}|\leq j_{\imath_2} \leq j_{\imath_1}+j_{e_3},\\\nonumber
&&...,|j_{\imath_{N_v-5}}-j_{e_{N_v-3}}|\leq j_{\imath_{N_v-4}}\leq j_{\imath_{N_v-5}}+j_{e_{N_v-3}}, \\\nonumber
&&|j_{\imath_{N_v-4}}-j_{e_{N_v-2}}|\leq j_{\imath_{N_v-3}}\leq j_{\imath_{N_v-4}}+j_{e_{N_v-2}},\\\nonumber
&&|j_{e_{N_v-1}}-j_{e_{N_v}}|\leq j_{\imath_{N_v-3}}\leq j_{e_{N_v-1}}+j_{e_{N_v}}.
\end{eqnarray}
Another basis of the  intertwiner space $\mathcal{H}_v^{\{j_e\}}$ is the so-called coherent intertwiner basis \cite{Livine:2007vk,long2019coherent}, which is established based on the $SU(2)$ coherent state.
A $SU(2)$ coherent state $| j, \hat{n} \rangle$ is defined via rotating the highest weight state $| j, j \rangle$ by $g(\hat{n}) \in SU(2)$, namely \cite{Perelomov:1986tf},
\begin{equation}
| j, \hat{n} \rangle = g(\hat{n}) | j , j \rangle.
\end{equation}
where $\hat{n} $ is a unit vector, and $g(\hat{n})\in SU(2)$ satisfies $\hat{n}= g(\hat{n}) \hat{z}$ with the north pole vector $\hat{z}=(0,0,1) \in \mathbb{S}^2$.
The $SU(2)$ coherent state $|j,\hat{n}\rangle$ can be decomposed by the orthonormal basis $\{|j,m\rangle\}$ as \cite{Perelomov:1986tf}
\begin{equation}
|j, \hat{n}\rangle=\sum_{m=-j}^jc_{j,m}(\hat{n})|j,m\rangle
\label{eq:CoherentState-UsualBasis}
\end{equation}
with,
\begin{equation}
c_{j,m}(\hat{n})=(\frac{(2j)!}{(j+m)!(j-m)!})^{\frac{1}{2}}(-\sin\frac{\pi-\theta}{2})^{j+m}(\cos\frac{\pi-\theta}{2})^{j-m} \exp(-\mathrm{i}(j+m)(\varphi+\pi)).
\end{equation}
Moreover, the $SU(2)$ coherent states $|j,\hat{n}\rangle$ provide a overcomplete basis of the space $V^j$ as
\begin{equation}
\mathbb{I}_{V^j}=(2j+1)\int_{\mathbb{S}^2}d n |j,\hat{n}\rangle\langle j,\hat{n}|,
\end{equation}
where $\mathbb{I}_{V^j}$ is the identity of $V^j$, and $dn$ is the normalized measure on the 2-sphere $S^2$.
Now, the coherent intertwiner  basis of $\mathcal{H}_v^{\{j_e\}}$ can be given as
\begin{equation}
\mathbb{I}_{\mathcal{H}_v^{\{j_e\}}}=\int_{\mathcal{S}_v^{\{j_e\}}}d\sigma_v^{\{j_e\}} |\mathcal{I}_{v,\{j_e\}}^{\{\hat{n}_e\}}\rangle\langle \mathcal{I}_{v,\{j_e\}}^{\{\hat{n}_e\}}|,
\end{equation}
where $\{\hat{n}_e\}\equiv(\hat{n}_{e_1}, \cdots ,\hat{n}_{e_{N_v}})$, $\mathcal{S}_v^{\{j_e\}}:=\{(\hat{n}_{e_1},\cdots,\hat{n}_{e_{N_v}})\in\times_{I=1}^{N_v}S_I^2|\sum_{e|t(e)=v}j_{e}\hat{n}_{e} -\sum_{e|s(e)=v}j_{e}\hat{n}_{e}=0\}/SU(2)$, $d\sigma_v^{\{j_e\}} $ is an invariant measure on $\mathcal{S}_v^{\{j_e\}}$, and the coherent intertwiner $|\mathcal{I}_{v,\{j_e\}}^{\{\hat{n}_e\}}\rangle$ is given by the following $SU(2)$-group-averaging
\begin{equation}
|\mathcal{I}_{v,\{j_e\}}^{\{\hat{n}_e\}}\rangle:=\int_{SU(2)}dg \bigotimes_{e|t(e)=v} g|j_{e},\hat{n}_{e}\rangle\bigotimes_{e|s(e)=v} \langle j_{e},\hat{n}_{e}|g^{-1}.
\end{equation}
In order to simplify our notations and distinguish the labels of the in-going and out-going edges, we use $j,\hat{n}$ to label the in-going edges and $\tilde{j},\hat{\tilde{n}}$ to the out-going edges. Then,
 the coherent intertwiner can be reformulated  as
\begin{equation}
|\mathcal{I}_{v,\{j,\tilde{j}\}}^{\{\hat{n},\hat{\tilde{n}}\}}\rangle:=\int_{SU(2)}dg \bigotimes_{I=1}^P g|j_I,\hat{n}_I\rangle\bigotimes_{J=1}^Q \langle \tilde{j}_{J},\hat{\tilde{n}}_{J}|g^{-1},
\end{equation}
where  $P$ is the number of the edges ended at $v$ and $Q$ is the number of the edges started at $v$.

The $SU(2)$ coherent states are said to be semi-classical states due to the property that they minimize the Heisenberg uncertainty relation  \cite{Perelomov:1986tf,Livine:2007vk,Long:2020euh}. A coherent spin state $| j, \hat{n} \rangle$ picks the unit vector $\hat{n}$ by $\vec{J}$ as $\hat{n}=\lim_{j \to \infty} \frac{\langle j, \hat{n} | \vec{J} | j , \hat{n} \rangle}{j}$. In the framework of LQG, each vertex $v \in \Gamma$ is dual to a polyhedron, and the edges attached to the $v$ are dual to the faces of the polyhedron \cite{PhysRevD.83.044035,Long:2020agv}. The area and the normal vector of the face are characterized by $j$ and $\hat{n}$ from $| j, \hat{n} \rangle$ respectively. These pair-data $\{ (j,\hat{n}) \}$ indeed provides a semi-classical but gauge variant picture . The gauge invariance is fulfilled via $SU(2)$-group-averaging over the tensoring spin states $\bigotimes_{e|t(e)=v} | j_{e} , \hat{n}_{e} \rangle\bigotimes_{e|s(e)=v} \langle j_{e} , \hat{n}_{e} |$ with $\{\hat{n}_e\}\in\mathcal{S}_v^{\{j_e\}}$, defining a gauge invariant coherent intertwiner. Although the information about the direction of each unit vector $\hat{n}_{e}$ loses due to the $SU(2)$-group-averaging, the relative angles amongst these unit vectors survive. Hence, a polyhedron in discrete geometry can be built from gauge invariant coherent intertwiners in a relational picture \cite{Livine:2007vk,PhysRevD.83.044035,Long:2020agv}: for a $v$ around by $N_v$ faces, the $\{ j_{e} \}$ determines $N_v$ areas and  $\{\hat{n}_e\}\in\mathcal{S}_v^{\{j_e\}}$ determines $2N_v-6$ relative angles.




\section{Density matrix and entanglement entropy of coherent intertwiner}


\subsection{Boundary and intertwiner entanglement entropies}\label{sec3.1}

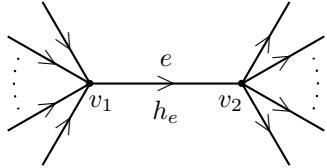
\begin{figure}[htb]
	\centering
	\begin{tikzpicture} [scale=1]

\coordinate  (P) at (0,0);
\coordinate  (Q) at (2,0);

\node[scale=0.7] at (P) {$\bullet$};
\node[scale=0.7] at (Q) {$\bullet$};

\draw (Q) ++ (240:0.3) node {$v_2$};

\draw[thick] (Q)  -- node[midway,sloped]{$>$} node[midway, above=0.1] {$e$} node[midway,below=0.1] {$h_e$} (P) ++ (-60:0.3) node {$v_1$};

\draw [thick, loosely dotted,domain=-20:20] plot ({2+1 * cos(\x)}, {0+1 * sin(\x)});

\draw [thick, loosely dotted,domain=160:200] plot ({1 * cos(\x)}, {1 * sin(\x)});

\draw[thick] (P)  to  node[midway,sloped]{$>$} ++ (120:1.25) ;

\draw[thick] (P)  to  node[midway,sloped]{$>$} ++ (150:1.25) ;

\draw[thick] (P)  to  node[midway,sloped]{$>$} ++ (240:1.25) ;

\draw[thick] (P)  to  node[midway,sloped]{$>$} ++ (210:1.25) ;

\draw[thick] (Q)  --  node[midway,sloped]{$>$} ++ (30:1.25) ;

\draw[thick] (Q)  to  node[midway,sloped]{$>$} ++ (60:1.25) ;

\draw[thick] (Q)  to  node[midway,sloped]{$>$} ++ (-30:1.25) ;

\draw[thick] (Q)  to  node[midway,sloped]{$>$} ++ (-60:1.25) ;

\end{tikzpicture}
\caption{The two-vertex graph $\Gamma$. The holomony $h_e$ along the edge $e$ that connects $v_1$ and $v_2$ contributes nothing to the entanglement entropy.
}
\label{fig:2-vertex}
\end{figure}

Consider a two-vertices graph $\Gamma$ as illustrated in Fig.\ref{fig:2-vertex}: one vertex $v_1$ is attached by $P+1$ edges and another vertex $v_2$ attached by $Q+1$ edges, meanwhile $v_1$ and $v_2$ are connected by edge $e$. For the sake of simplifying the notations, we re-orient all of the edges to ensure that, $e$ is out-going at $v_1$ and in-going at $v_2$ with other edges being in-going at $v_1$ and out-going at $v_2$ without losing generality.
Then, the boundary Hilbert space for this system is defined as below
\begin{equation}
\mathcal{H}_{\partial \Gamma}:=\bigotimes_{I=1}^{P} V^{I} \otimes \bigotimes_{J=1}^{Q} \bar{V}^{J}.
\end{equation}
That is, excluding the Hilbert space associated with the internal edge $e$ whose two-ends are contained in the graph. The apparent bipartition on the boundary Hilbert space is given by
\begin{equation}
\mathcal{H}_{\partial \Gamma}
=
\mathcal{H}_{v_{1}}^{\partial} \otimes \mathcal{H}_{v_{2}}^{\partial},
\quad \text{where} \
\mathcal{H}_{v_{1}}^{\partial}:=\bigotimes_{I=1}^{P} V^{I},
\ \text{and} \
\mathcal{H}_{v_{2}}^{\partial}:=\bigotimes_{J=1}^{Q} \bar{V}^{J}.
\end{equation}
On the other hand, the intertwiner Hilbert space on graph $\Gamma$ is defined as below
\begin{equation}
\mathcal{H}_{\Gamma}:=\mathcal{H}_{v_1} \otimes \mathcal{H}_{v_2},
\quad \text{where} \
\mathcal{H}_{v_{1}}:=\mathrm{Inv} \left( \bigotimes_{I=1}^{P} V_{I} \otimes \bar{V}_{e} \right)
\ \text{and} \
\mathcal{H}_{v_{2}}:=\mathrm{Inv} \left( V^{e} \otimes \bigotimes_{J=1}^{Q} \bar{V}_{J} \right).
\end{equation}
The gluing map \cite{Livine:2017fgq} provides a correspondence between the boundary state $| \psi_{\Gamma} \rangle_{\partial} \in \mathcal{H}_{\partial \Gamma}$ and the intertwiner state $| \psi_{\Gamma} \rangle \in \mathcal{H}_{\Gamma}$. Intuitively, the gluing map glues the boundary edges together along the internal edges.
Let us illustrate the gluing map as follows. Assume that the internal edge $e$ that links $v_1$ and $v_2$ carries a fixed $j$. Then, the spin-network state is written as
\begin{eqnarray}
| \Psi_{\Gamma}^{j} \rangle=
&&\sum_{ \{ j_{I}, \tilde{j}_{J},j_{\imath}, \tilde{j}_{\imath} \} } \mathcal{C}_{\mathcal{I}^{\{ j_{I} \}, j }_{v_1,\{j_{\imath}\}} \mathcal{I}^{\{ \tilde{j}_{J} \}, j }_{v_2,\{\tilde{j}_{\imath}\}} }
\underbrace{
\left(
\sum_{ m, \{ m_{I} \} }
\mathcal{I}^{ j_{1} \cdots j_{P} j,\{j_{\imath}\} }_{m_{1} \cdots m_{P} m } \bigotimes_{I=1}^{P} | j_{I}, m_{I} \rangle \langle j, m |
\right)}_{= | \mathcal{I}^{\{ j_{I} \}, j }_{v_1,\{j_{\imath}\}} \rangle}\\\nonumber
&&\qquad \qquad \qquad\otimes
\underbrace{
\left(
\sum_{ n, \{ \tilde{m}_{J} \} }
| j, n \rangle \bigotimes_{J=1}^{Q} \langle \tilde{j}_{J}, \tilde{m}_{J} | \mathcal{I}^{ j \tilde{j}_{1} \cdots \tilde{j}_{Q},\{\tilde{j}_{\imath}\} }_{ n \tilde{m}_{1} \cdots \tilde{m}_{Q} }
\right)}_{= | \mathcal{I}^{\{ \tilde{j}_{J} \}, j }_{v_2} \rangle}.
\label{eq:Generic2verticesSpinNet}
\end{eqnarray}
Let us explain the notations in this equation: (i) $ j_{I}, \tilde{j}_{J}$ label the boundary edges of $\Gamma$ in left and right side respectively, and $j_{\imath}, \tilde{j}_{\imath}$ represent the recoupling spins for the coherent intertwiner basis $\mathcal{I}^{\{ j_{I} \}, j }_{v_1,\{j_{\imath}\}}$ and $  \mathcal{I}^{\{ \tilde{j}_{J} \}, j }_{v_2,\{\tilde{j}_{\imath}\}} $ respectively; (ii) The coefficients $\mathcal{I}^{ j_{1} \cdots j_{P} j,\{j_{\imath}\} }_{m_{1} \cdots m_{P} m }$ as well as $\mathcal{I}^{ j \tilde{j}_{1} \cdots \tilde{j}_{Q},\{\tilde{j}_{\imath}\} }_{ n \tilde{m}_{1} \cdots \tilde{m}_{Q} }$ are responsible for the  gauge invariances of  $\mathcal{I}^{\{ j_{I} \}, j }_{v_1,\{j_{\imath}\}}$ and $  \mathcal{I}^{\{ \tilde{j}_{J} \}, j }_{v_2,\{\tilde{j}_{\imath}\}} $  respectively, which  can be constructed by concatenating Clebsch-Gordan coefficients; (iii) The coefficients $\mathcal{C}_{\mathcal{I}^{\{ j_{I} \}, j }_{v_1,\{j_{\imath}\}} \mathcal{I}^{\{ \tilde{j}_{J} \}, j }_{v_2,\{\tilde{j}_{\imath}\}} }$ encode the correlation between intertwiners living at $v_1$ and $v_2$.
On the other hand, the boundary state associated with the spin-network is written as
\begin{eqnarray}\label{gluing}
| \Psi_{\Gamma}^{j} \rangle_{\partial}
=&&
\sum_{ \{ j_{I}, \tilde{j}_{J},j_{\imath}, \tilde{j}_{\imath} \} }\sqrt{2j+1} \mathcal{C}_{\mathcal{I}^{\{ j_{I} \}, j }_{v_1,\{j_{\imath}\}} \mathcal{I}^{\{ \tilde{j}_{J} \}, j }_{v_2,\{\tilde{j}_{\imath}\}} }\\\nonumber
&&\cdot\sum_{ m,n,\{ m_{I}, \tilde{m}_{J} \} }
\mathcal{I}^{ j_{1} \cdots j_{P} j ,\{j_{\imath}\}}_{m_{1} \cdots m_{P} m } \bigotimes_{I=1}^{P} | j_{I}, m_{I} \rangle  D^{j}_{mn}(h_e) \bigotimes_{J=1}^{Q} \langle \tilde{j}_{J}, \tilde{m}_{J} | \mathcal{I}^{ j \tilde{j}_{1} \cdots \tilde{j}_{Q} ,\{\tilde{j}_{\imath}\}}_{ n \tilde{m}_{1} \cdots \tilde{m}_{Q} },
\end{eqnarray}
where $D^{j}_{mn}(h_e):=\langle j,m|h_e|j,n\rangle$. The gluing map is then viewed as sending $| \Psi_{\Gamma}^{j} \rangle$ to $| \Psi_{\Gamma}^{j} \rangle_{\partial}$, via sandwiching holonomy that associates the edge to-be-glued. This gluing map can be also established without the holonomy insertion (i.e., setting $h_e=\text{identity}$ in Eq. \eqref{gluing}), which leads to 
\begin{equation}
|\check{ \Psi}_{\Gamma}^{j} \rangle_{\partial}
=
\sum_{ \{ j_{I}, \tilde{j}_{J},j_{\imath}, \tilde{j}_{\imath} \} }\sqrt{2j+1} \mathcal{C}_{\mathcal{I}^{\{ j_{I} \}, j }_{v_1,\{j_{\imath}\}} \mathcal{I}^{\{ \tilde{j}_{J} \}, j }_{v_2,\{\tilde{j}_{\imath}\}} }
\sum_{ m,\{ m_{I}, \tilde{m}_{J} \} }
\mathcal{I}^{ j_{1} \cdots j_{P} j,\{j_{\imath}\} }_{m_{1} \cdots m_{P} m } \bigotimes_{I=1}^{P} | j_{I}, m_{I} \rangle \bigotimes_{J=1}^{Q} \langle \tilde{j}_{J}, \tilde{m}_{J} | \mathcal{I}^{ j \tilde{j}_{1} \cdots \tilde{j}_{Q},\{\tilde{j}_{\imath}\} }_{ m \tilde{m}_{1} \cdots \tilde{m}_{Q} }.
\end{equation}

Now we look at the entanglement carried by these states.
The intertwiner entanglement entropy $E(v_1 | v_2 )$ with respect to $\mathcal{H}_{\Gamma}=\mathcal{H}_{v_{1}} \otimes \mathcal{H}_{v_{2}}$ is given by the Von Neumann entropy from below reduced density matrix
\begin{align}
\rho^j_{v_1}
&:=
\text{Tr}_{\mathcal{H}_{v_{2}}}
| \Psi_{\Gamma}^{j} \rangle \langle \Psi_{\Gamma}^{j} |
\\
&=
\sum_{ \{ j_{I}, j'_{I},j_{\imath},j'_{\imath} \} } \sum_{ \{ \tilde{j}_{J},\tilde{j}_{\imath} \} } \mathcal{C}_{\mathcal{I}^{\{ j_{I} \}, j }_{v_1,\{j_{\imath}\}} \mathcal{I}^{\{ \tilde{j}_{J} \}, j }_{v_2,\{\tilde{j}_{\imath}\}} } \overline{\mathcal{C}_{\mathcal{I}^{\{ j'_{I} \}, j }_{v_1,\{j'_{\imath}\}} \mathcal{I}^{\{ \tilde{j}_{J} \}, j }_{v_2,\{\tilde{j}_{\imath}\}} }}
| \mathcal{I}^{\{ j_{I} \}, j }_{v_1,\{j_{\imath}\}} \rangle \langle \mathcal{I}^{\{ j'_{I} \}, j }_{v_1,\{j'_{\imath}\}} |.
\end{align}
On the other hand, the boundary entanglement entropy $E(v^{\partial}_1 | v^{\partial}_2 )$ with respect to $\mathcal{H}_{\partial \Gamma}=\mathcal{H}_{v_{1}}^{\partial} \otimes \mathcal{H}_{v_{2}}^{\partial}$ is given by the Von Neumann entropy from below reduced density matrix
\begin{align}
&
\rho_{v_1}^{{\scriptscriptstyle  \partial},j}
:=
\text{Tr}_{\mathcal{H}_{v_{2}}^{\partial}}
| \Psi_{\Gamma}^{j} \rangle \langle \Psi_{\Gamma}^{j} |_{\partial}=\text{Tr}_{\mathcal{H}_{v_{2}}^{\partial}}
| \check{\Psi}_{\Gamma}^{j} \rangle \langle \check{\Psi}_{\Gamma}^{j} |_{\partial}
\\
=&
\sum_{ \{ j_{I}, j'_{I},j_{\imath},j'_{\imath} \} } \sum_{ \{ \tilde{j}_{J},\tilde{j}_{\imath} \} } \mathcal{C}_{\mathcal{I}^{\{ j_{I} \}, j }_{v_1,\{j_{\imath}\}} \mathcal{I}^{\{ \tilde{j}_{J} \}, j }_{v_2,\{\tilde{j}_{\imath}\}} } \overline{\mathcal{C}_{\mathcal{I}^{\{ j'_{I} \}, j }_{v_1,\{j'_{\imath}\}} \mathcal{I}^{\{ \tilde{j}_{J} \}, j }_{v_2,\{\tilde{j}_{\imath}\}} }}
\sum_{ m, \{ m_{I}, m'_{I} \} }
\mathcal{I}^{ j_{1} \cdots j_{P} j,\{j_{\imath}\} }_{m_{1} \cdots m_{P} m } \overline{\mathcal{I}^{ j'_{1} \cdots j'_{P} j ,\{j'_{\imath}\}}_{m'_{1} \cdots m'_{P} m }} \bigotimes_{I=1}^{P} | j_{I}, m_{I} \rangle\langle j'_{I}, m'_{I} |
\nonumber
\\
=&
\sum_{ \{ j_{I}, j'_{I},j_{\imath},j'_{\imath} \} } \sum_{ \{ \tilde{j}_{J},\tilde{j}_{\imath} \} } \mathcal{C}_{\mathcal{I}^{\{ j_{I} \}, j }_{v_1,\{j_{\imath}\}} \mathcal{I}^{\{ \tilde{j}_{J} \}, j }_{v_2,\{\tilde{j}_{\imath}\}} } \overline{\mathcal{C}_{\mathcal{I}^{\{ j'_{I} \}, j }_{v_1,\{j'_{\imath}\}} \mathcal{I}^{\{ \tilde{j}_{J} \}, j }_{v_2,\{\tilde{j}_{\imath}\}} }}
\sum_{ m }
\langle j, m | \mathcal{I}^{\{ j_{I} \}, j }_{v_1,\{j_{\imath}\}} \rangle \langle \mathcal{I}^{\{ j'_{I} \}, j }_{v_1,\{j'_{\imath}\}} | j, m \rangle,
\end{align}
where $\mathcal{H}_{v_{2}}^{\partial}=\bigotimes_{J=1}^{Q}\bar{V}_{J}$ is the boundary-edge state space attached to the vertex $v_2$, and below orthogonality is used
\begin{equation}
\sum_{\{ m_{I} \}_{I=1}^{N} }
\mathcal{I}^{ j_{1} \cdots j_{N} j,\{j_{\imath}\} }_{m_{1} \cdots m_{N} m } \overline{\mathcal{I}^{ j_{1} \cdots j_{N} j',\{j'_{\imath}\} }_{m_{1} \cdots m_{N} m' }}
=
\frac{1}{2j+1}
\delta_{jj'}\delta_{mm'}\delta_{\{j_{\imath}\}\{j'_{\imath}\}}.
\end{equation}
Indeed, one can verify the normalization $\text{Tr}_{\mathcal{H}_{v_{1}}^{\partial}} \rho_{v_1}^{{\scriptscriptstyle \partial},j}=1$.
Note that the relation between the two density matrices can be given by
\begin{align}
\rho_{v_1}^{{\scriptscriptstyle  \partial},j}
=\text{Tr}_{V_{e}} \rho^j_{v_1}
\label{eq:ReducedDensityMatrices-Fixedj}
\end{align}
with $V_e=V^{j}$.
That is to say, the reduced density matrix  $\rho_{v_1}^{{\scriptscriptstyle \partial},j}$ of the half-cut boundary can be understood as tracing the reduced density matrix $ \rho^j_{v_1}$ of the half-cut graph over the recoupled Hilbert space (here, it is $V^j$ associative with the spin-$j$ along the internal edge $e$).
Then, following \cite{Livine:2017fgq} one can show a simple relation between the entanglement entropy of $\mathcal{H}_{\partial \Gamma}=\mathcal{H}_{v_{1}}^{\partial} \otimes \mathcal{H}_{v_{2}}^{\partial}$ and of $\mathcal{H}_{\Gamma}=\mathcal{H}_{v_{1}} \otimes \mathcal{H}_{v_{2}}$.
\begin{theorem} \label{thm:EntanglementEntropies-FixedSpin}
In the cases that the spin along the internal edge is fixed at $j$, the following relation between entanglement entropies holds
\begin{equation}\label{EEln}
E_j(v^{\partial}_1 | v^{\partial}_2 )=E_j(v_1 | v_2) + \ln (2j+1),
\end{equation}
where $E_j(v^{\partial}_1 | v^{\partial}_2 ):=-\text{Tr}(\rho_{v_1}^{{\scriptscriptstyle \partial},j}\ln\rho_{v_1}^{{\scriptscriptstyle \partial},j})$ and $E_j(v_1 | v_2 ):=-\text{Tr}(\rho^j_{v_1}\ln\rho^j_{v_1})$ .
\end{theorem}
The relation Eq.(\ref{EEln}) between boundary entanglement and intertwiner entanglement can be generalized straightforwardly to the cases that the internal edge $e$ carries spin-superposition. More explicitly,
one can consider the state 
\begin{equation}\label{superpositionj}
| \Psi_{\Gamma} \rangle
=
\sum_{j} \alpha_{j} | \Psi_{\Gamma}^{j} \rangle
=
\sum_{j} \sum_{ \{ j_{I}, \tilde{j}_{J},j_{\imath}, \tilde{j}_{\imath} \} }\alpha_{j} \mathcal{C}_{\mathcal{I}^{\{ j_{I} \}, j }_{v_1,\{j_{\imath}\}} \mathcal{I}^{\{ \tilde{j}_{J} \}, j }_{v_2,\{\tilde{j}_{\imath}\}} }
| \mathcal{I}^{\{ j_{I} \}, j }_{v_1,\{j_{\imath}\}} \rangle
\otimes
| \mathcal{I}^{\{ \tilde{j}_{J} \}, j }_{v_2,\{\tilde{j}_{\imath}\}} \rangle,
\end{equation}
 the gluing state with holonomy insertion 
\begin{eqnarray}\label{gluing2}
| \Psi_{\Gamma} \rangle_{\partial}
&=&
\sum_{j} \alpha_{j}\sum_{ \{ j_{I}, \tilde{j}_{J},j_{\imath}, \tilde{j}_{\imath} \} }\sqrt{2j+1} \mathcal{C}_{\mathcal{I}^{\{ j_{I} \}, j }_{v_1,\{j_{\imath}\}} \mathcal{I}^{\{ \tilde{j}_{J} \}, j }_{v_2,\{\tilde{j}_{\imath}\}} }\\\nonumber
&&\cdot\sum_{ m,n,\{ m_{I}, \tilde{m}_{J} \} }
\mathcal{I}^{ j_{1} \cdots j_{P} j,\{j_{\imath}\} }_{m_{1} \cdots m_{P} m } \bigotimes_{I=1}^{P} | j_{I}, m_{I} \rangle  D^{j}_{mn}(h_e) \bigotimes_{J=1}^{Q} \langle \tilde{j}_{J}, \tilde{m}_{J} | \mathcal{I}^{ j \tilde{j}_{1} \cdots \tilde{j}_{Q} ,\{\tilde{j}_{\imath}\}}_{ n \tilde{m}_{1} \cdots \tilde{m}_{Q} },
\end{eqnarray}
and the gluing state without holonomy insertion
\begin{eqnarray}
&&| \check{\Psi}_{\Gamma} \rangle_{\partial}\\\nonumber
&=&
\sum_{j} \alpha_{j}\sum_{ \{ j_{I}, \tilde{j}_{J},j_{\imath}, \tilde{j}_{\imath} \} }\sqrt{2j+1} \mathcal{C}_{\mathcal{I}^{\{ j_{I} \}, j }_{v_1,\{j_{\imath}\}} \mathcal{I}^{\{ \tilde{j}_{J} \}, j }_{v_2,\{\tilde{j}_{\imath}\}} }
\sum_{ m,\{ m_{I}, \tilde{m}_{J} \} }
\mathcal{I}^{ j_{1} \cdots j_{P} j,\{j_{\imath}\} }_{m_{1} \cdots m_{P} m } \bigotimes_{I=1}^{P} | j_{I}, m_{I} \rangle   \bigotimes_{J=1}^{Q} \langle \tilde{j}_{J}, \tilde{m}_{J} | \mathcal{I}^{ j \tilde{j}_{1} \cdots \tilde{j}_{Q},\{\tilde{j}_{\imath}\} }_{ m \tilde{m}_{1} \cdots \tilde{m}_{Q} }.
\end{eqnarray}
Again, the reduced density matrices are obtained via partial tracing in $\mathcal{H}_{v_{2}}^{\partial}$ and $\mathcal{H}_{v_{2}}$ respectively, which gives,
\begin{equation}\label{traceequal}
\tilde{\rho}_{v_1}^{\partial}
:=
\text{Tr}_{\mathcal{H}_{v_{2}}^{\partial}}
| \Psi_{\Gamma}\rangle \langle \Psi_{\Gamma} |_{\partial}=\text{Tr}_{\mathcal{H}_{v_{2}}^{\partial}}
| \check{\Psi}_{\Gamma}\rangle \langle \check{\Psi}_{\Gamma} |_{\partial},
\end{equation}
and 
\begin{equation}
\tilde{\rho}_{v_1}:=\text{Tr}_{\mathcal{H}_{v_{2}}}
| \Psi_{\Gamma} \rangle \langle \Psi_{\Gamma} |.
\end{equation}
Similarly, the generalized relation between $\tilde{\rho}_{v_1}^{\partial}$ and $\tilde{\rho}_{v_1}$ holds by taking superposition of $j$ into account, which reads
\begin{align}
\tilde{\rho}_{v_1}^{\partial}
=\text{Tr}_{V_{e}} \tilde{\rho}_{v_1},
\end{align}
where $V_e=\bigoplus_{j}V^j$ and $\tilde{\rho}_{v_1}=\bigoplus_{j}\tilde{p}_j\rho^j_{v_1}$ with $\tilde{p}_{j}\equiv\alpha_{j}\bar{\alpha}_j$. Then, the relation between intertwiners and boundary  entanglements leads the following Theorem.
\begin{theorem} \label{thm:EntanglementEntropies-SuperposedSpins}
In the cases that the spin along the internal edge is superposed, say the intertwiner state is given by $| \Psi_{\Gamma} \rangle$,
then the following relation between entanglement entropies holds
\begin{equation}\label{3terms}
E(v^{\partial}_1 | v^{\partial}_2 )=\sum_{j} \tilde{p}_{j} \ln (2j+1) - \sum_{j} \tilde{p}_{j} \ln \tilde{p}_{j} + \sum_{j} \tilde{p}_{j}E_j(v_1 | v_2),
\end{equation}
where $E(v^{\partial}_1 | v^{\partial}_2 ):=-\text{Tr}(\tilde{\rho}_{v_1}^{\partial}\ln\tilde{\rho}_{v_1}^{\partial})$ and  $E_j(v_1 | v_2 )=-\text{Tr}(\rho^j_{v_1}\ln\rho^j_{v_1})$.
\end{theorem}

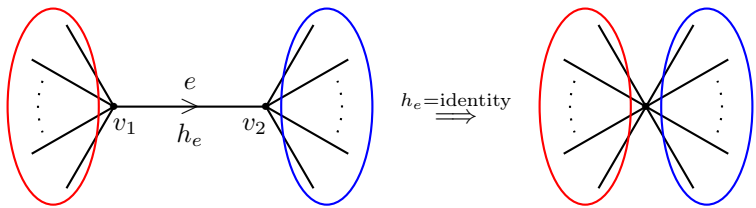
\begin{figure}[htb]
	\centering
	\begin{tikzpicture} [scale=1]

\coordinate  (P) at (0,0);
\coordinate  (Q) at (2,0);

\node[scale=0.7] at (P) {$\bullet$};
\node[scale=0.7] at (Q) {$\bullet$};

\draw (Q) ++ (240:0.3) node {$v_2$};

\draw[thick] (Q)  -- node[midway,sloped]{$>$} node[midway, above=0.1] {$e$} node[midway,below=0.1] {$h_e$} (P) ++ (-60:0.3) node {$v_1$};

\draw [thick, loosely dotted,domain=-20:20] plot ({2+1 * cos(\x)}, {0+1 * sin(\x)});

\draw [thick, loosely dotted,domain=160:200] plot ({1 * cos(\x)}, {1 * sin(\x)});

\draw[thick] (P)  to ++ (120:1.25) ;

\draw[thick] (P)  to ++ (150:1.25) ;

\draw[thick] (P)  to ++ (240:1.25) ;

\draw[thick] (P)  to ++ (210:1.25) ;

\draw[thick] (Q)  to ++ (30:1.25) ;

\draw[thick] (Q)  to ++ (60:1.25) ;

\draw[thick] (Q)  to ++ (-30:1.25) ;

\draw[thick] (Q)  to ++ (-60:1.25) ;

\draw[thick,red] (P) ++ (-0.8,0) ellipse (0.6 and 1.3);
\draw[thick,blue] (Q) ++ (0.8,0) ellipse (0.6 and 1.3);

\draw (Q) ++ (2.5,0) node {$\overset{h_e=\text{identity}}{\Longrightarrow}$};
\coordinate  (O) at (7,0);

\node[scale=0.7] at (O) {$\bullet$};

\draw [thick, loosely dotted,domain=-20:20] plot ({7+1 * cos(\x)}, {0+1 * sin(\x)});

\draw [thick, loosely dotted,domain=160:200] plot ({7+1 * cos(\x)}, {1 * sin(\x)});

\draw[thick] (O)  to ++ (120:1.25) ;

\draw[thick] (O)  to ++ (150:1.25) ;

\draw[thick] (O)  to ++ (240:1.25) ;

\draw[thick] (O)  to ++ (210:1.25) ;

\draw[thick] (O)  to ++ (30:1.25) ;

\draw[thick] (O)  to ++ (60:1.25) ;

\draw[thick] (O)  to ++ (-30:1.25) ;

\draw[thick] (O)  to ++ (-60:1.25) ;

\draw[thick,red] (O) ++ (-0.8,0) ellipse (0.6 and 1.3);
\draw[thick,blue] (O) ++ (0.8,0) ellipse (0.6 and 1.3);

\end{tikzpicture}
\caption{The left graph (two-vertex) has the same entanglement entropy between boundary edges with the right graph (one-vertex).
}
\label{fig:2-vertex-1-vertex}
\end{figure}

Let us have a discussion on the above Theorem.  First, referring to \cite{Livine:2017fgq}, the first term in Eq.\eqref{3terms} should be interpreted as coming from gauge-breaking,
and it follows that the second and the third terms should be interpreted as the intertwiner entanglement, since the $- \sum_{j} \tilde{p}_{j} \ln \tilde{p}_{j}$ comes from the spin-superposition along the linking edge $e$, and the $E_j(v_1 | v_2)$ is the intertwiner entanglement when the spin is fixed. 
Second, in the case of single internal edge graph, it has been shown that the holonomy along the internal edge $e$ plays no role in the entanglement entropy in Ref.\cite{Livine:2017fgq} for fixed $j$, and Eq.\eqref{traceequal} in our calculation extends this point to the case of superposed $j$; Indeed, one is able to gauge-fix the $h_e$ into identity, and then $| {\Psi}_{\Gamma} \rangle_{\partial}$ becomes $| \check{\Psi}_{\Gamma} \rangle_{\partial}$ which can be regarded as an intertwiner on a single vertex as illustrated  in Fig. \ref{fig:2-vertex-1-vertex};  Notice that  Eq.\eqref{traceequal} tells us that $| {\Psi}_{\Gamma} \rangle_{\partial}$ and $| \check{\Psi}_{\Gamma} \rangle_{\partial}$ have the same reduced density matrix, thus the respect entanglement entropies depicted in Fig. \ref{fig:2-vertex-1-vertex} are indistinguishable. 
Third, while above general formalism is still vague for the sake of establishing relation between the entanglement and geometry,  the coherent intertwiners provide a semi-classical picture of geometry on polyhedron, which could interlace the genuine quantum notion --- entanglement, with the discrete geometry.
In the following part of this paper, we are going to explore how the entanglement entropy emerges from this semi-classical picture, and how the entanglement gets reflected in the discrete geometry, or in turn.

\subsection{Entanglement produced from group-averaging}\label{sec3.2}

As a prelude for the study on coherent intertwiner, this part is meant to show how entanglement can be produced from group averaging.
We begin with a gauge variant scenario based on the graph with only one  vertex $v$, and the corresponding wave-function is below tensor state,
\begin{align}
\bigotimes_{e|t(e)=v} | j_{e} , m_{e} \rangle \otimes \bigotimes_{e|s(e')=v} \langle j_{e'} , m_{e'} |
\in
\bigotimes_{e|t(e)=v} V^{j_{e}} \otimes \bigotimes_{e|s(e')=v} \bar{V}^{j_{e'}}.
\end{align}
This is not a physical spin-network due to the absence of gauge invariance, and there is also no entanglement. 
To get a gauge invariant state, the group-averaging is adopted, which inevitably introduces superposition and entanglement. It is possible to grant the physical implication for the group-averaging by considering some $SU(2)$ invariant measurement: suppose that we are given a set $\{ | \phi_{i} \rangle \}_{i}$ whose members are all $SU(2)$-invariant pointer-states $| \phi_{i} \rangle$, i.e., $| \phi_{i} \rangle = g | \phi_{i} \rangle$ for any $g \in SU(2)$, then this invariance can be conveyed to the probability distribution $| \langle \phi_{i} | \psi \rangle |^2$ where $| \psi \rangle$ is the state to be observed, because
\begin{align}
\langle \phi_{i} | \psi \rangle
=
\int_{SU(2)} dg \langle g \phi_{i} | \psi \rangle
=
\int_{SU(2)} dg \langle \phi_{i} | g^{\dagger} \psi \rangle
=
\langle \phi_{i} | \int_{SU(2)} dg | g^{\dagger} \psi \rangle.
\end{align}
It is clear that $\int_{SU(2)} dg | g^{\dagger} \psi \rangle$ is $SU(2)$-invariant. This group-averaging process can be viewed as a particular `twirling', say $\rho \mapsto \int_{SU(2)} dg (g \rho g^{\dagger})$ in the field of quantum information, with slightly different that the group-average here is based on a pure state $| \psi \rangle$ and getting another pure state $\int_{SU(2)} dg | g^{\dagger} \psi \rangle$, while twirling can be implemented on any state, and usually getting a mixed state.

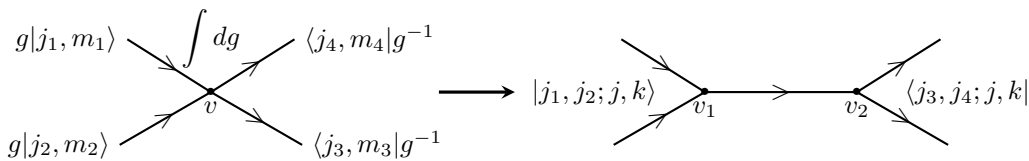
\begin{figure}[htb]
	\centering
	\begin{tikzpicture} [scale=1]
\coordinate  (O) at (0,0);

\coordinate  (P) at (6.5,0);
\coordinate  (Q) at (8.5,0);

\node[scale=0.7] at (O) {$\bullet$} node[below] {$v$}node [above=0.2] {$\displaystyle{ \int d g }$};

\node[scale=0.7] at (P) {$\bullet$};
\node[scale=0.7] at (Q) {$\bullet$};

\draw[thick] (O)  --  node[midway,sloped]{$>$} ++ (1.1,0.7) node[right] {$\langle j_4, m_4 | g^{-1} $};

\draw[thick] (O)  to  node[midway,sloped]{$>$} ++ (1.2,-0.7) node[right] {$\langle j_3, m_3 | g^{-1} $};

\draw[thick] (O)  to  node[midway,sloped]{$>$} ++ (-1.2,-0.7) node[left] {$g | j_2, m_2 \rangle$};

\draw[thick] (O)  to  node[midway,sloped]{$>$} ++ (-1.1,0.7) node[left] {$g | j_1, m_1 \rangle$};

\draw[->,>=stealth,very thick] (O) ++ (3,0) to  ++ (1,0);

\draw[thick] (Q) node[below] {$v_2$} node [right=0.5] {$\langle j_3 , j_4 ; j ,k|$} -- node[midway,sloped]{$>$} (P) node[below] {$v_1$} node [left=0.5] {$| j_1 , j_2 ; j ,k\rangle$};

\draw[thick] (Q)  --  node[midway,sloped]{$>$} ++ (1.1,0.7) ;

\draw[thick] (Q)  to  node[midway,sloped]{$>$} ++ (1.2,-0.7) ;

\draw[thick] (P)  to  node[midway,sloped]{$>$} ++ (-1.2,-0.7) ;

\draw[thick] (P)  to  node[midway,sloped]{$>$} ++ (-1.1,0.7) ;

\end{tikzpicture}
\caption{The $SU(2)$-group-averaging produces entanglement between the legs. This can be viewed as the entanglement between two coupled states $| j_1 , j_2 ; j ,k\rangle$ and $\langle j_3 , j_4 ; j ,k |$ at $v_1$ and $v_2$, respectively.
}
\label{fig:4valent-GroupAveraging}
\end{figure}

The rest of this part will calculate the entanglement introduced by the group-averaging. Let us look at an example. Consider the graph with only one four-valents' vertex $v$ and the intertwiner space $\check{\mathcal{H}}_v^{\{j_1,j_2,j_3,j_4\}}= V^{j_1}\otimes V^{j_2}\otimes \bar{V}^{j_3}\otimes \bar{V}^{j_4}$ on $v$. An element in $\check{\mathcal{H}}_v^{\{j_1,j_2,j_3,j_4\}}$ is given by the tensor product as below,
\begin{align}
| j_{1}, m_{1} \rangle \otimes | j_{2}, m_{2} \rangle \otimes \langle j_{3}, m_{3} | \otimes \langle j_{4}, m_{4} |,
\end{align}
where we take the four-valents' bipartition $(2,2)$, with the two left-side edges being ingoing and the two right-side edges being outgoing. 
Clearly, there is no entanglement between left and right sides, say $E(A|B)=0$ with $\mathcal{H}_A:=V^{j_1}\otimes V^{j_2}$ and $\mathcal{H}_B:= \bar{V}^{j_3}\otimes \bar{V}^{j_4}$. Now, let us implement the $SU(2)$-group-averaging over the tensor state, which leads
\begin{align}\label{eq:GroupAveraging-Magnetics}
| \check{\mathcal{I}} \rangle
&=
\int_{SU(2)} dg
g | j_{1}, m_{1} \rangle \otimes g | j_{2}, m_{2} \rangle \otimes \langle j_{3}, m_{3} | g^{-1} \otimes \langle j_{4}, m_{4} | g^{-1}
\\\nonumber
&=
\sum_{j} \sum_{k,m=-j}^{j} \sum_{ \vec{k} }
\frac{1}{2j+1} C^{j_1 j_2 j}_{k_1 k_2 k} \overline{C^{j_1 j_2 j}_{m_1 m_2 m} } \overline{C^{j_3 j_4 j}_{k_3 k_4 k}} C^{j_3 j_4 j}_{m_3 m_4 m}
| j_{1}, k_{1} \rangle \otimes | j_{2}, k_{2} \rangle \otimes \langle j_{3}, k_{3} | \otimes \langle j_{4}, k_{4} |,
\end{align}
where $\vec{k}\equiv\{k_1,k_2,k_3,k_4\}$ , $C^{j_1 j_2 j}_{m_1 m_2 m}\equiv\langle j_1,m_1,j_2,m_2 | j, m \rangle$ stands for Clebsch-Gordan coefficient. One should note that the group-averaging spoils the normalization so it should be retrieved by rescaling later.
In addition, recall that $\vec{m}\equiv\{m_1,m_2,m_3,m_4\} $ are fixed, the $SU(2)$-group-averaging will eliminate some configuration that do not satisfy $m_1+m_2=m_3+m_4$. This is the closure condition on magnetic quantum numbers.
The state $| \check{\mathcal{I}} \rangle$ survived from the group-averaging is a gauge invariant state, and it can be also viewed as a bipartite system between two sets of recoupled spins, see the illustration in Fig.\ref{fig:4valent-GroupAveraging}. To simplify the expression, let us rewrite $|\check{\mathcal{I}}\rangle$ as  
\begin{align}
|\check{\mathcal{I}}\rangle
&=
\sum_{j,k}
\frac{1}{2j+1} f(j,\vec{j},\vec{m})
| j_1, j_2 ; j,k \rangle \otimes \langle j_3, j_4 ; j,k |
 ,
\end{align}
where $| j_1, j_2 ; j,k \rangle := \sum_{ \{k _1,k_2 \} }C^{j_1 j_2 j}_{k_1 k_2 k}  | j_{1}, k_{1} \rangle \otimes | j_{2}, k_{2} \rangle$ defines a recoupled spin, likewise for $\langle j_3, j_4 ; j,k |$, and  we denote
\begin{align}
f(j,\vec{j},\vec{m})\equiv\sum_{m} \overline{C^{j_1 j_2 j}_{m_1 m_2 m}} C^{j_3 j_4 j}_{m_3 m_4 m}
\label{eq:amplitude-magnetics}
\end{align}
 for the fixed  $\vec{j}\equiv\{j_1,j_2,j_3,j_4\} $ and $\vec{m}$.

\begin{figure}[htbp]
    \centering 
    \subfloat[$j_1=j_2=j_3=j_4=5$]{\label{j=5}\includegraphics[width=0.5\linewidth]{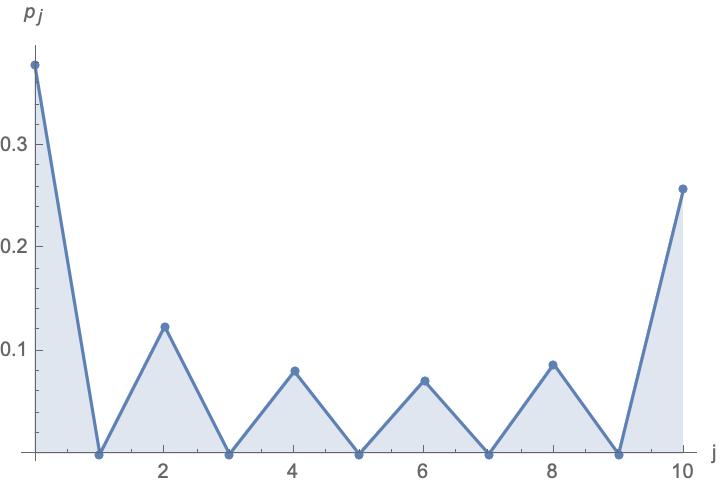}}  
    \subfloat[$j_1=j_2=j_3=j_4=10$]{\label{j=10}\includegraphics[width=0.5\linewidth]{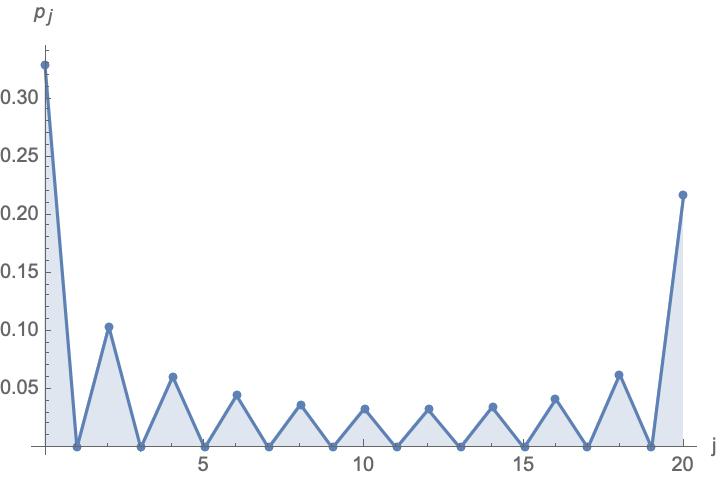}}  
    \\    
    \subfloat[$j_1=j_2=j_3=j_4=20$]{\label{j=25}\includegraphics[width=0.5\linewidth]{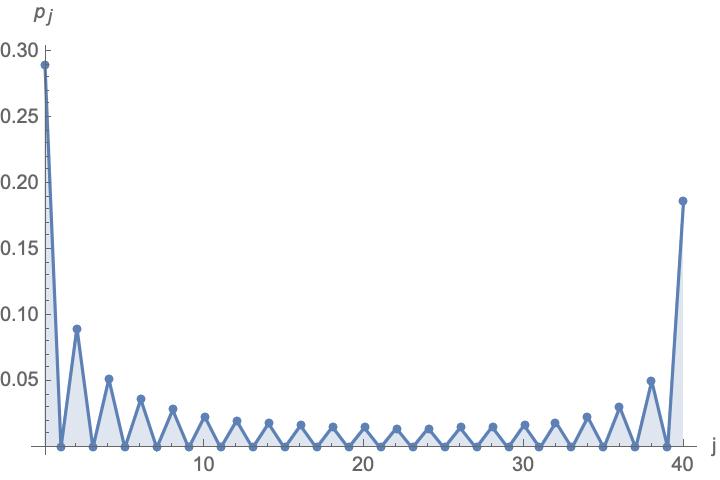}}  
    \subfloat[$j_1=j_2=j_3=j_4=30$]{\label{j=50}\includegraphics[width=0.5\linewidth]{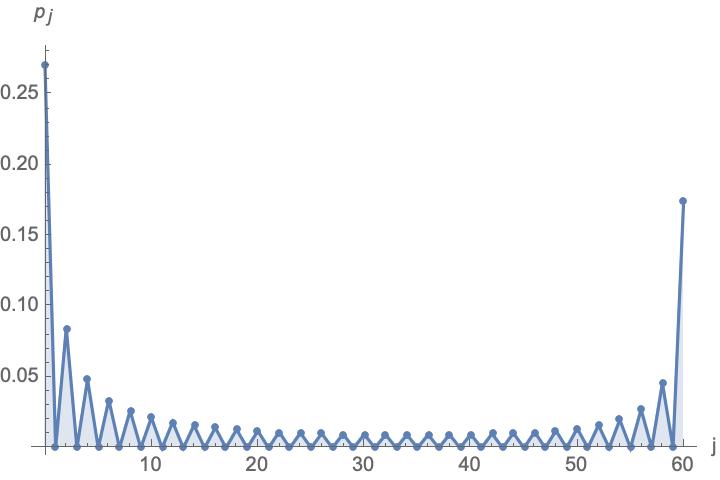}}   
     \caption{The numerical results of $p_j$ for $(m_1,m_2,m_3,m_4)=(0,0,0,0)$ and $j_1=j_2=j_3=j_4=5,10,20,30$, where the $x$-axis shows the recouping spin $j$ and $y$-axis shows the numerical value of $p_j$. These figures show that $p_j$ has an oscillation with respect to coupling spin $j$.}  \label{m000}
\end{figure}

\begin{figure}[htbp]
    \centering 
    \subfloat[$j_1=j_2=j_3=j_4=5$]{\label{j=5}\includegraphics[width=0.5\linewidth]{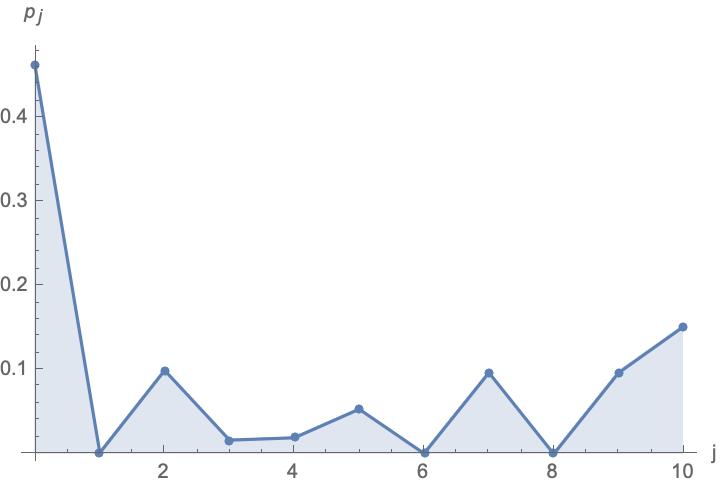}}  
    \subfloat[$j_1=j_2=j_3=j_4=10$]{\label{j=10}\includegraphics[width=0.5\linewidth]{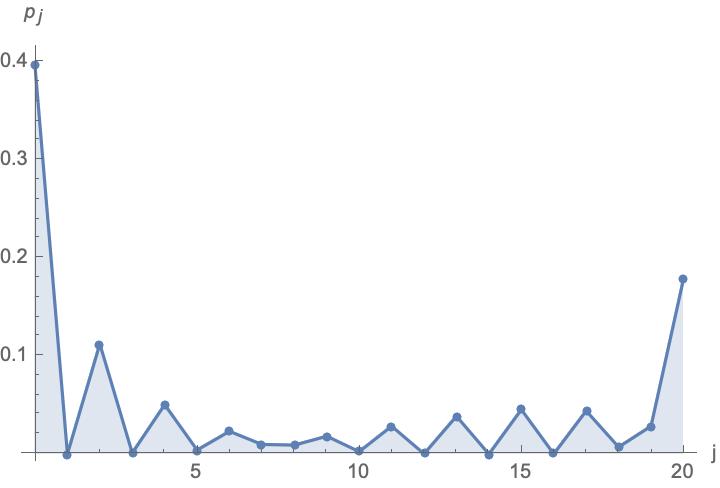}}  
    \\    
    \subfloat[$j_1=j_2=j_3=j_4=20$]{\label{j=25}\includegraphics[width=0.5\linewidth]{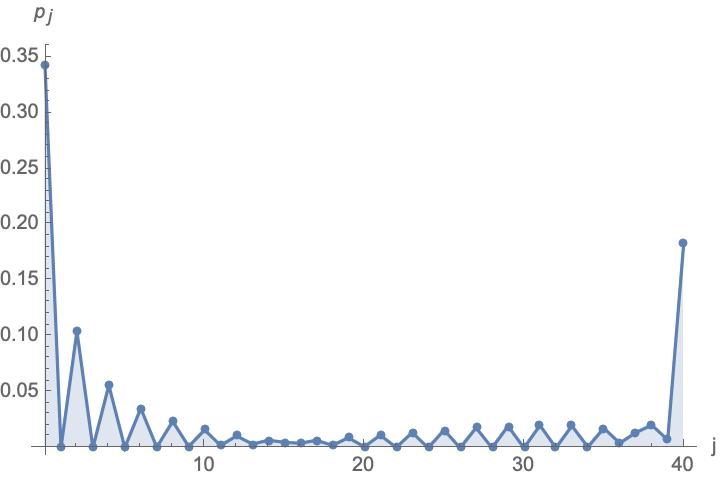}}  
    \subfloat[$j_1=j_2=j_3=j_4=30$]{\label{j=50}\includegraphics[width=0.5\linewidth]{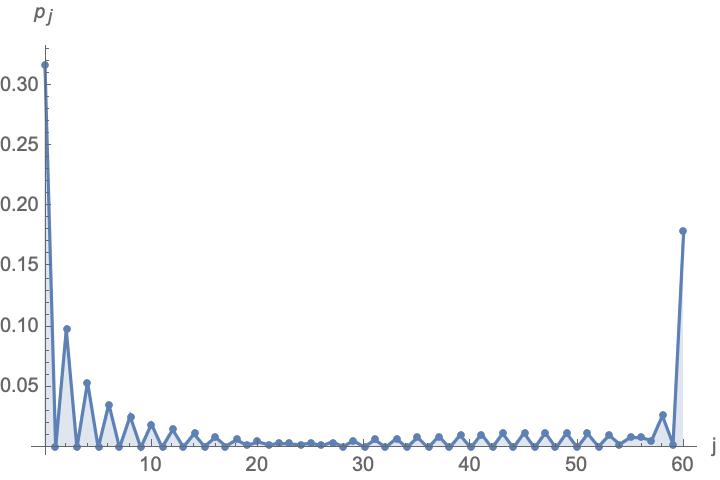}}   
     \caption{The numerical results of $p_j$ for $(m_1,m_2,m_3,m_4)=(1,-1,1,-1)$ and $j_1=j_2=j_3=j_4=5,10,20,30$, where the $x$-axis shows the recouping spin $j$ and $y$-axis shows the numerical value of $p_j$. These figures show the difference in the shape of oscillation compared to the cases of $m_1=m_2=m_3=m_4=0$.} \label{m111}
\end{figure}

\begin{figure}[htbp]
    \centering 
    \subfloat[$j_1=j_2=j_3=j_4=5$]{\label{j=5}\includegraphics[width=0.5\linewidth]{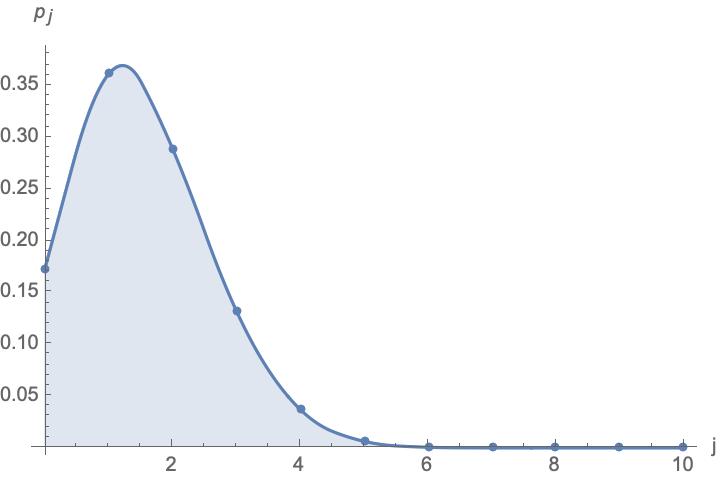}}  
    \subfloat[$j_1=j_2=j_3=j_4=10$]{\label{j=10}\includegraphics[width=0.5\linewidth]{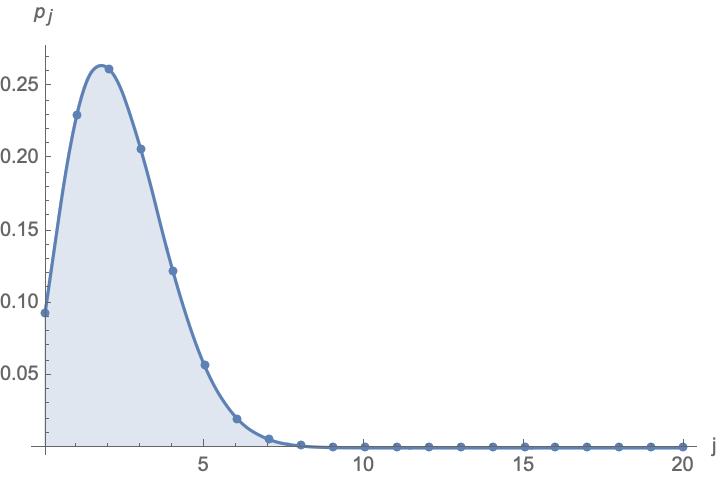}}  
    \\    
    \subfloat[$j_1=j_2=j_3=j_4=20$]{\label{j=25}\includegraphics[width=0.5\linewidth]{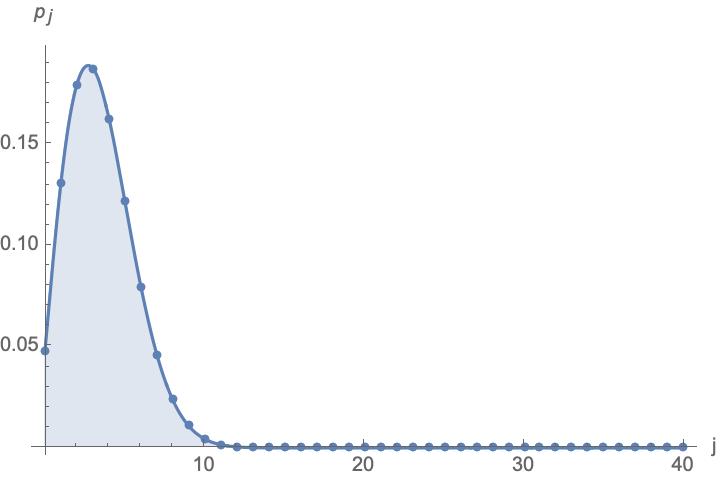}}  
    \subfloat[$j_1=j_2=j_3=j_4=30$]{\label{j=50}\includegraphics[width=0.5\linewidth]{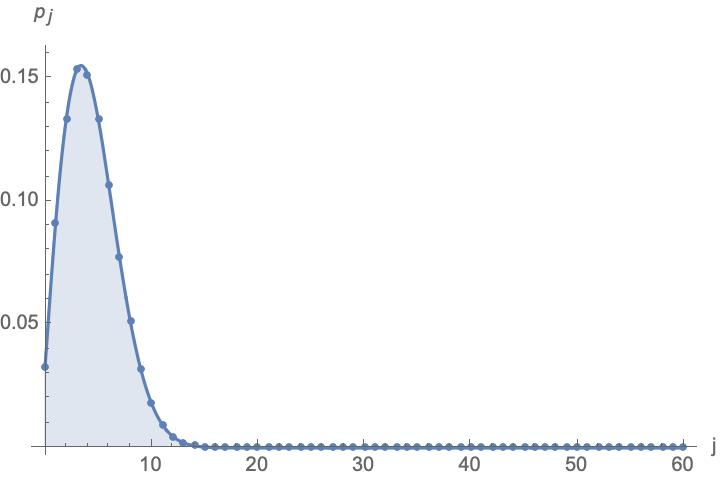}}   
     \caption{The numerical results of $p_j$ for $j_1=j_2=j_3=j_4=5,10,20,30$ and $m_1=m_3=-m_2=-m_4=j_4$, where the $x$-axis shows the recoupling spin $j$ and $y$-axis shows the numerical value of $p_j$. These figures show that the shape of the distributions $p_j$ are peaks for the highest (and lowest) weight states.} \label{m222}
\end{figure}

Recall the bipartition $\mathcal{H}_A:=V^{j_1}\otimes V^{j_2}$ and $\mathcal{H}_B:= \bar{V}^{j_3}\otimes \bar{V}^{j_4}$. Then, the entanglement $E(A|B)$ between $A$ and $B$ can be given by the Von Neumann entropy of the reduced density matrices $\rho_{A}$.
For the state $|\check{\mathcal{I}}\rangle$, the reduced density matrix ${\rho}_{A}$ is defined by ${\rho}_{A}:=\text{Tr}_B(\rho_{\check{\mathcal{I}}})$ with $\rho_{\check{\mathcal{I}}}\equiv\frac{|\check{\mathcal{I}}\rangle\langle\check{\mathcal{I}}|}{\langle\check{\mathcal{I}}|\check{\mathcal{I}}\rangle}$. More explicitly, one has
\begin{align}
\rho_{A}
=
\frac{1}{\langle \check{\mathcal{I}} | \check{\mathcal{I}} \rangle}
\sum_{j} \sum_{k=-j}^{j} \frac{ | f(j,\vec{j},\vec{m}) |^2 }{(2j+1)^2} | j_1, j_2 ; j,k \rangle \langle j_1, j_2 ; j,k |
.
\end{align}
One can introduce the probability distribution $p_j$ of the recoupling spin $j$, which is given by $p_j=\frac{| f(j,\vec{j},\vec{m}) |^2}{ (2j+1)\langle \check{\mathcal{I}} | \check{\mathcal{I}} \rangle}$, and the reduced density matrix can be decomposed into ${\rho}^{A}_{j} $, i.e.,
\begin{align}
{\rho}_{A}
=
\sum_{j} p_{j} {\rho}^{A}_{j},
\quad {\rho}^{A}_{j}=\sum_{k=-j}^{j} \frac{| j_1, j_2 ; j,k \rangle \langle j_1, j_2 ; j,k |}{(2j+1)}
.
\end{align}
It is clear that the $\mathcal{H}_{A}$ and $\mathcal{H}_{B}$ are entangled for the state $| \check{\mathcal{I}} \rangle$. The entanglement entropy $E(A|B):=-\text{Tr}(\rho_{A}\ln \rho_{A})$ is determined by the distribution $p_j$, which reads
\begin{align}
E(A|B)=E_p(A|B)+E_0(A|B),\quad E_p(A|B):=-\sum_{j}(p_j\ln p_j),\quad E_0(A|B):=\sum_{j}p_j S^A_j,
\end{align}
where $S^A_j:=-\text{Tr}(\rho^{A}_j\ln \rho_j^{A})=\ln(2j+1)$. Further, the distribution $p_j$ and the entanglement entropy $E(A|B)$ can be calculated numerically. The numerical results of $p_j$ are illustrated in Figs.\ref{m000},\ref{m111} and \ref{m222}, which show that the distribution $p_j$ is oscillating with respect to $j$ for small $\{m_1,m_2,m_3,m_4\}$ state, while there is a peak for highest (and lowest) weight state.
The numerical values of entanglement entropy $E(A|B)$ are listed in Tabs. \ref{tab00} and  \ref{tab0}, which show that the entanglement entropy can be controlled by  the magnetic configurations.

It is worth to have a discussion on these results. First, one should notice that the quantum number $\{m_1,m_2,m_3,m_4\}$ are gauge variant, and their geometric interpretation become fuzzy after group-averaging. Second, note that the distribution $p_j$ is a peak for highest (and lowest) weight state, it  ensures that the entanglement entropy is able to capture  the main character of the distribution $p_j$. By combining these two points, it is reasonable to consider the entanglement carried by gauge invariant coherent intertwiners, since they are constructed by the highest (and lowest) weight state and they describe  semiclassical geometry on polyhedrons.
In next subsection, we will focus on the coherent intertwiners which provide a semiclassical picture of polyhedron geometry, and one may expect that both entanglement, superposition, and the geometric picture could be drawn by the gauge invariant knowledge encoded in the area-weighted normal vectors $\{j_e\hat{n}_{e} \}$ labelling the coherent intertwiners. 
%
\begin{table}
    \centering
    \begin{tabular}{|c|c|c|c|c|l} \hline  
         $(j_1,j_2,j_3,j_4)$&  $(m_1,m_2,m_3,m_4)$ &  $E_p(A|B)$ & $E_0(A|B)$ & $E(A|B)$ \\ \hline  
         $(5,5,5,5)$&  $(0,0,0,0)$ &  1.58138 & 1.5931 & 3.17448 \\ \hline  
         $(5,5,5,5)$& $(1,-1,1,-1)$ & 1.64349 & 1.37831 & 3.0218 \\ \hline  
         $(5,5,5,5)$& $(2,-2,2,-2)$ & 1.68047 & 1.35532 & 3.03579 \\ \hline  
         $(5,5,5,5)$& $(3,-3,3,-3)$ &  1.72176 & 1.33854 & 3.06029 \\ \hline 
 $(5,5,5,5)$& $(4,-4,4,-4)$ & 1.6484 & 1.30671& 2.95511 \\\hline 
 $(5,5,5,5)$& $(5,-5,5,-5)$ & 1.45701 & 1.21724 & 2.67426\\ \hline
 $(5,5,5,5)$& $(2,-1,2,-1)$ & 1.7005 & 2.05912 & 3.75962 \\ \hline
 $(5,5,5,5)$& $(3,-1,3,-1)$ & 1.77649 & 2.32768 & 4.10417 \\ \hline
 $(5,5,5,5)$& $(4,-1,4,-1)$ & 1.59752 & 2.48947 & 4.08698 \\ \hline
 $(5,5,5,5)$& $(5,-1,5,-1)$ & 1.31356 & 2.60331 & 3.91687 \\ \hline
    \end{tabular}
    \caption{The numerical values of $E(A|B)$ for small spins $j_1=j_2=j_3=j_4=5$ at different configurations with respect to magnetic numbers.}\label{tab00}
\end{table} 

\begin{table}
    \centering
    \begin{tabular}{|c|c|c|c|c|l} \hline  
         $(j_1,j_2,j_3,j_4)$&  $(m_1,m_2,m_3,m_4)$ &  $E_p(A|B)$ & $E_0(A|B)$ & $E(A|B)$ \\ \hline  
         $(10,10,10,10)$&  $(0,0,0,0)$ &  2.01747 & 2.02953 & 4.047 \\ \hline  
         $(10,10,10,10)$& $(1,-1,1,-1)$ & 2.01367 & 1.80432 & 3.81799 \\ \hline  
         $(10,10,10,10)$& $(2,-2,2,-2)$ & 2.06684 & 1.77726 & 3.8441 \\ \hline  
         $(10,10,10,10)$& $(3,-3,3,-3)$ &  2.10459 & 1.76694 & 3.87153 \\ \hline 
 $(10,10,10,10)$& $(4,-4,4,-4)$ & 2.1037 & 1.76034 & 3.86404 \\\hline 
 $(10,10,10,10)$& $(5,-5,5,-5)$ & 2.13296 & 1.75342 & 3.88638 \\ \hline
 $(10,10,10,10)$& $(6,-6,6,-6)$ & 2.16708 & 1.7438 & 3.91088 \\ \hline
 $(10,10,10,10)$& $(7,-7,7,-7)$ & 2.1329 & 1.72895 & 3.86185 \\ \hline
 $(10,10,10,10)$& $(8,-8,8,-8)$ & 2.08743 & 1.70467 & 3.7921 \\ \hline
 $(10,10,10,10)$& $(9,-9,9,-9)$ & 1.99391 & 1.66086 & 3.65477 \\ \hline
  $(10,10,10,10)$& $(10,-10,10,-10)$ & 1.77857 & 1.55688 & 3.33545 \\ \hline
    \end{tabular}
    \caption{The numerical values of $E(A|B)$ for small spins $j_1=j_2=j_3=j_4=10$ at different configurations with respect to magnetic numbers.}\label{tab0}
\end{table}







\subsection{Entanglement entropy  between legs of coherent intertwiner}\label{3.2}



\subsubsection{Coherent intertwiner with 4-legs}\label{sec3.3.1}
Let us consider the gauge invariant coherent intertwiner space $\mathcal{H}_v^{\{j_1,j_2,j_3,j_4\}}=\text{Inv}_{SU(2)}[ V^{j_1}\otimes V^{j_2}\otimes \bar{V}^{j_3}\otimes \bar{V}^{j_4}]$  on a four-valents vertex, in which an element is given by
\begin{equation}
|\mathcal{I}\rangle=\int_{SU(2)}dg g|j_1,\hat{n}_1\rangle\otimes g|j_2,\hat{n}_2\rangle\otimes\langle j_3,\hat{n}_3|g^{-1}\otimes\langle j_4,\hat{n}_4|g^{-1}.
\end{equation}
\begin{figure}[htb]
	\centering
	\begin{tikzpicture} [scale=1]
\coordinate  (O) at (0,0);

\coordinate  (P) at (6.5,0);
\coordinate  (Q) at (8.5,0);

\node[scale=0.7] at (O) {$\bullet$} node[below] {$v$}node [above=0.2] {$\displaystyle{ \int d g }$};

\node[scale=0.7] at (P) {$\bullet$};
\node[scale=0.7] at (Q) {$\bullet$};

\draw[thick] (O)  --  node[midway,sloped]{$>$} ++ (1.1,0.7) node[right] {$\langle j_4, \hat{n}_4 | g^{-1} $};

\draw[thick] (O)  to  node[midway,sloped]{$>$} ++ (1.2,-0.7) node[right] {$\langle j_3, \hat{n}_3 | g^{-1} $};

\draw[thick] (O)  to  node[midway,sloped]{$>$} ++ (-1.2,-0.7) node[left] {$g | j_2, \hat{n}_2 \rangle$};

\draw[thick] (O)  to  node[midway,sloped]{$>$} ++ (-1.1,0.7) node[left] {$g | j_1, \hat{n}_1 \rangle$};

\draw[->,>=stealth,very thick] (O) ++ (3,0) to  ++ (1,0);

\draw[thick] (Q) node[below] {$v_2$} node [right=0.5] {$\langle j_3 , j_4 ; j,k |$} --node[midway,sloped]{$>$} (P) node[below] {$v_1$} node [left=0.5] {$| j_1 , j_2 ; j ,k\rangle$};

\draw[thick] (Q)  --  node[midway,sloped]{$>$} ++ (1.1,0.7) ;

\draw[thick] (Q)  to  node[midway,sloped]{$>$} ++ (1.2,-0.7) ;

\draw[thick] (P)  to  node[midway,sloped]{$>$} ++ (-1.2,-0.7) ;

\draw[thick] (P)  to  node[midway,sloped]{$>$} ++ (-1.1,0.7) ;

\end{tikzpicture}
\caption{The $SU(2)$-group-averaging on spin coherent state  produces entanglement between the legs. This can be viewed as the entanglement between two coupled states $| j_1 , j_2 ; j,k \rangle$ and $\langle j_3 , j_4 ; j ,k |$ at $v_1$ and $v_2$, respectively.
}
\label{fig:4valent-CS-GroupAverage}
\end{figure}
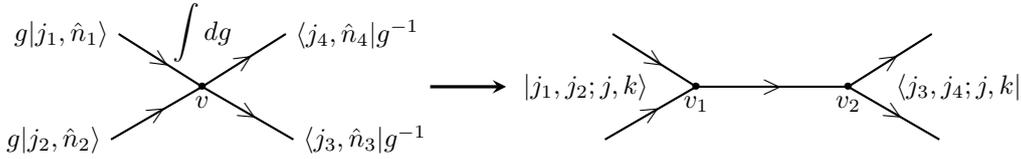
By recalling Eq.\eqref{eq:CoherentState-UsualBasis}, the state is then rewritten in the manner of repeating Eq.\eqref{eq:GroupAveraging-Magnetics} as illustrated in Fig.\ref{fig:4valent-CS-GroupAverage}, which reads
\begin{align}
\ket{ \mathcal{I} }
&=\sum_{\vec{m}} c_{j_1,m_1}(\hat{n}_1) c_{j_2,m_2}(\hat{n}_2) \bar{c}_{j_3,m_3}(\hat{n}_3) \bar{c}_{j_4,m_4}(\hat{n}_4)\\\nonumber
&\quad \cdot\int_{SU(2)}dg g \ket{j_1  ,m_1}\otimes\ket{j_2, m_2 }\otimes \bra{ j_3 , m_3}\otimes\bra{j_4 , m_4 }g^{-1}
\nonumber
\\
&=
\sum_{\vec{m}} c_{j_1,m_1}(\hat{n}_1) c_{j_2,m_2}(\hat{n}_2) \bar{c}_{j_3,m_3}(\hat{n}_3) \bar{c}_{j_4,m_4}(\hat{n}_4)
\sum_{j,m,k} \frac{ \overline{C^{j_1 j_2 j}_{m_1 m_2 m}} C^{j_3 j_4 j}_{m_3 m_4 m} }{2j+1} \ket{ j_1,j_2;j,k} \otimes \bra{ j_3,j_4;j,k},
\nonumber \\
&=\sum_{j,m,k} \frac{ C^{j_1 j_2 j}_{\hat{n}_1 \hat{n}_2 m} \overline{ C^{j_3 j_4 j}_{\hat{n}_3 \hat{n}_4 m} } }{2j+1} \ket{ j_1,j_2;j,k} \otimes \bra{ j_3,j_4;j,k}
\end{align}
where the second line uses the recoupled spins $\ket{j_1,j_2;j,k}:=\sum_{m_1,m_2} C^{j_1j_2j}_{m_1 m_2 k}|j_1,m_1\rangle \otimes|j_2,m_2\rangle$ and $\bra{ j_3,j_4;j,k}:=\sum_{m_3,m_4} \overline{C^{j_3j_3j}_{m_3 m_4 k}}\langle j_3,m_3|\otimes\langle j_4,m_4|$, and in the third line we have denoted
\begin{align}
C^{j_1 j_2 j}_{\hat{n}_1 \hat{n}_2 m}
&\equiv\sum_{m_1=-j_1}^{j_1} \sum_{m_2=-j_2}^{j_2}  c_{j_1,m_1}(\hat{n}_1) c_{j_2,m_2}(\hat{n}_2)
\overline{C^{j_1 j_2 j}_{m_1 m_2 m}},
\\
\overline{ C^{j_3 j_4 j}_{\hat{n}_3 \hat{n}_4 m} }
&\equiv\sum_{m_3=-j_3}^{j_3} \sum_{m_4=-j_4}^{j_4}  \bar{c}_{j_3,m_3}(\hat{n}_3) \bar{c}_{j_4,m_4}(\hat{n}_4)
C^{j_3 j_4 j}_{m_3 m_4 m}.
\end{align}
The density matrix of the coherent intertwiner $|\mathcal{I}\rangle$ is given by $\rho_{\mathcal{I}}\equiv\frac{|\mathcal{I}\rangle\langle \mathcal{I}|}{\langle \mathcal{I}|\mathcal{I}\rangle}$ where its denominator, namely, the normalization factor, is given by
\begin{eqnarray}
\langle \mathcal{I}|\mathcal{I}\rangle
=
\sum_{j} \frac{1}{2j+1}
\left| \sum_{m} C^{j_1 j_2 j}_{\hat{n}_1 \hat{n}_2 m} \overline{ C^{j_3 j_4 j}_{\hat{n}_3 \hat{n}_4 m} } \right|^2.
\end{eqnarray}

Recall the definitions $\mathcal{H}_A:=V^{j_1}\otimes V^{j_2}$ and $\mathcal{H}_B:= \bar{V}^{j_3}\otimes \bar{V}^{j_4}$. Then, for the state $|\mathcal{I}\rangle$, the entanglement $E(A|B)$ between $A$ and $B$ can be given by the Von Neumann entropy of the reduced density matrices $\rho_{A}$, which are defined by
\begin{equation}
\rho_{A}:=\text{Tr}_B(\rho_{\mathcal{I}}),\quad E(A|B):=-\text{Tr}(\rho_{A}\ln \rho_{A}).
\end{equation}
More explicitly, $\rho_{A}$ can be calculated as
\begin{eqnarray}
\rho_{A}&=&\text{Tr}_B(\rho_{\mathcal{I}})\\\nonumber
   &=& \sum_{j} \frac{1}{2j+1}\frac{ \left| \sum_{m} C^{j_1 j_2 j}_{\hat{n}_1 \hat{n}_2 m} \overline{ C^{j_3 j_4 j}_{\hat{n}_3 \hat{n}_4 m} } \right|^2  }{\sum_{j'} \frac{1}{2j'+1}\left| \sum_{ m'} C^{j_1 j_2 j'}_{\hat{n}_1 \hat{n}_2 m'} \overline{ C^{j_3 j_4 j'}_{\hat{n}_3 \hat{n}_4 m'} } \right|^2 } \sum_{k=-j}^{j} \frac{ | j_1,j_2;j,k\rangle\langle j_1,j_2;j,k | }{ 2j+1 },
\end{eqnarray}
which can be also read in the form of
\begin{align}
\rho_{A}=\sum_{j} p_{j} \rho_j^{A}, \qquad
p_{j} =\frac{ \left| \sum_{m} C^{j_1 j_2 j}_{\hat{n}_1 \hat{n}_2 m} \overline{ C^{j_3 j_4 j}_{\hat{n}_3 \hat{n}_4 m} } \right|^2 }
{ (2j+1) \left| \sum_{j', m} \frac{1}{2j'+1}C^{j_1 j_2 j'}_{\hat{n}_1 \hat{n}_2 m''} \overline{ C^{j_3 j_4 j'}_{\hat{n}_3 \hat{n}_4 m'} } \right|^2 },
\end{align}
and $\rho_j^{A}=\frac{\sum_{k} |j_1,j_2;j,k\rangle\langle j_1,j_2;j,k| }{(2j+1)}$.  Further, we have
\begin{equation}
E(A|B)=E_p(A|B)+E_0(A|B),\quad E_p(A|B):=-\sum_{j}(p_j\ln p_j),\quad E_0(A|B):=\sum_{j}p_j S^A_j
\end{equation}
with $S^A_j\equiv-\text{Tr}(\rho^{A}_j\ln \rho_j^{A})=\ln(2j+1)$.

\begin{figure}[htbp]
    \centering 
    \subfloat[$j_1=j_2=j_3=j_4=5$]{\label{j=5}\includegraphics[width=0.5\linewidth]{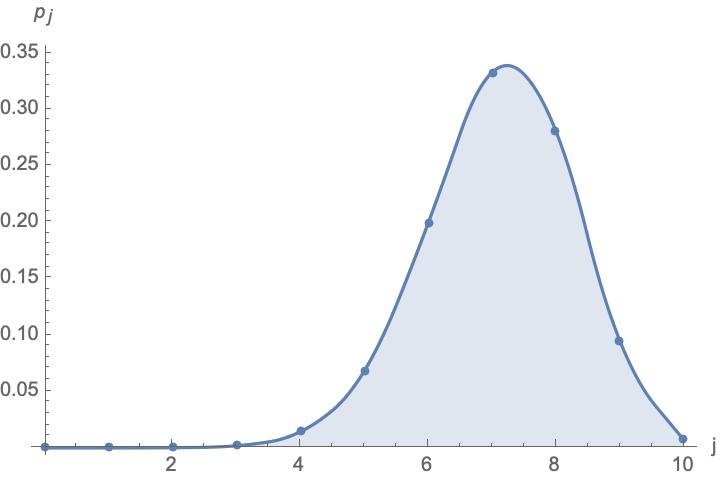}}  
    \subfloat[$j_1=j_2=j_3=j_4=10$]{\label{j=10}\includegraphics[width=0.5\linewidth]{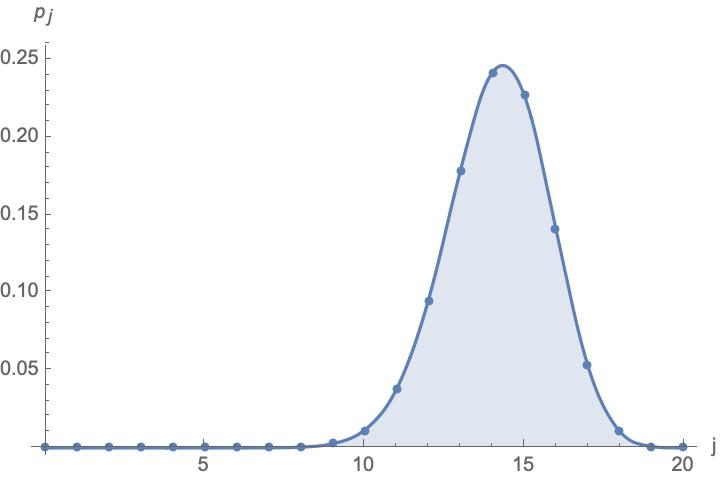}}  
    \\    
    \subfloat[$j_1=j_2=j_3=j_4=20$]{\label{j=20}\includegraphics[width=0.5\linewidth]{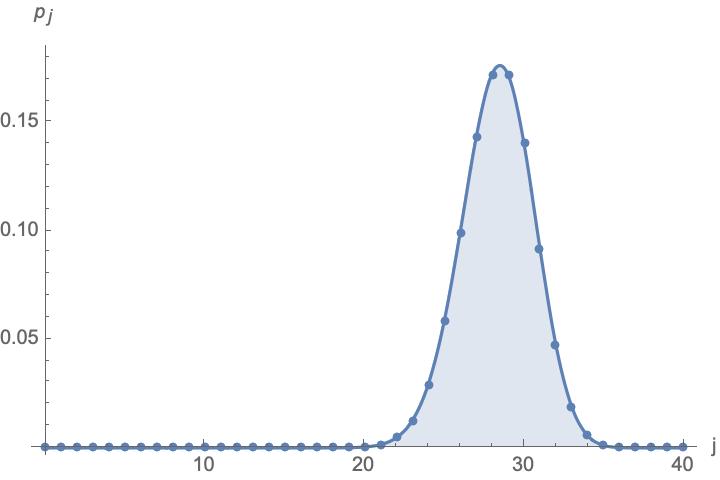}}  
    \subfloat[$j_1=j_2=j_3=j_4=30$]{\label{j=30}\includegraphics[width=0.5\linewidth]{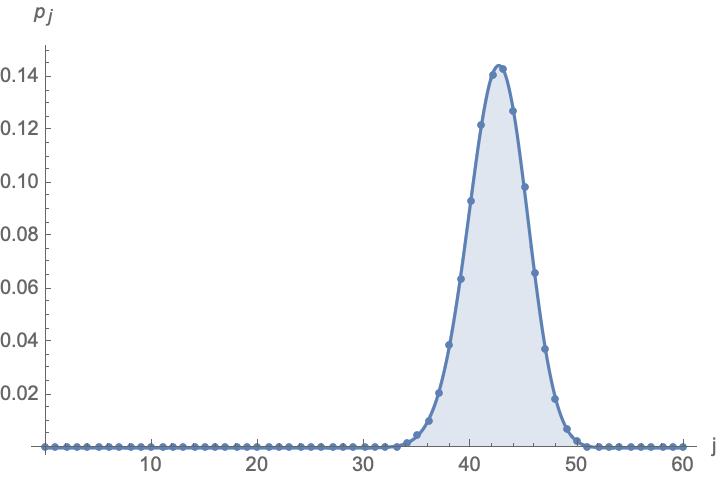}}   
     \caption{The numerical results of $p_j$ for $(\theta_1,\theta_2,\theta_3,\theta_4)=(\frac{\pi}{2},\frac{\pi}{2},\frac{\pi}{4},\frac{3\pi}{4}),(\varphi_1,\varphi_2,\varphi_3,\varphi_4)=(0,\frac{\pi}{2},\frac{\pi}{4},\frac{\pi}{4})$ and $j_1=j_2=j_3=j_4=5,10,20,30$, where the $x$-axis shows the recoupling spin $j$ and $y$-axis shows the numerical value of $p_j$. These figures show that $p_j$ has a peak near $j=j_0\equiv|j_1\hat{n}_1+j_2\hat{n}_2|$, and this peak shrinks relative to the  range of $j$ with the boundary spins $j_1,j_2,j_3,j_4$ getting larger.}
     \label{fig1}    
\end{figure}

\begin{figure}[htbp]
    \centering 
    \subfloat[ $\vec{\theta}=(\frac{\pi}{2},\frac{\pi}{2},\frac{7\pi}{16},\frac{9\pi}{16}),  \vec{\varphi}=(0,\frac{\pi}{8},\frac{\pi}{16},\frac{\pi}{16})$]
    {\label{1d8}\includegraphics[width=0.5\linewidth]{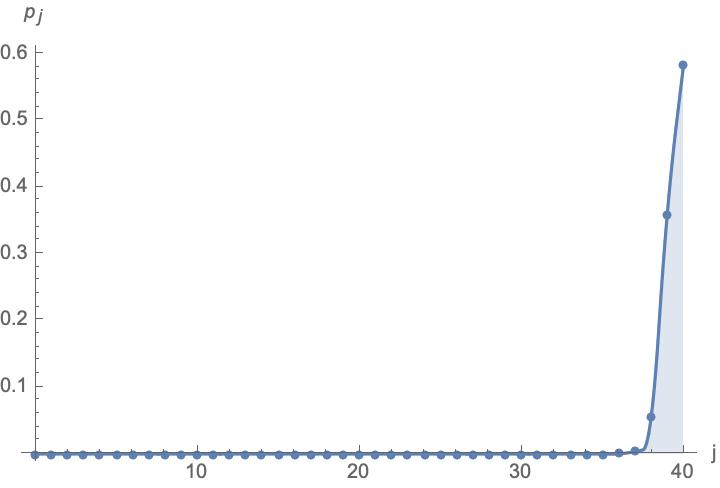}}  
    \subfloat[$\vec{\theta}=(\frac{\pi}{2},\frac{\pi}{2},\frac{3\pi}{8},\frac{5\pi}{8}), \vec{\varphi}=(0,\frac{\pi}{4},\frac{\pi}{8},\frac{\pi}{8})$]{\label{1d4}\includegraphics[width=0.5\linewidth]{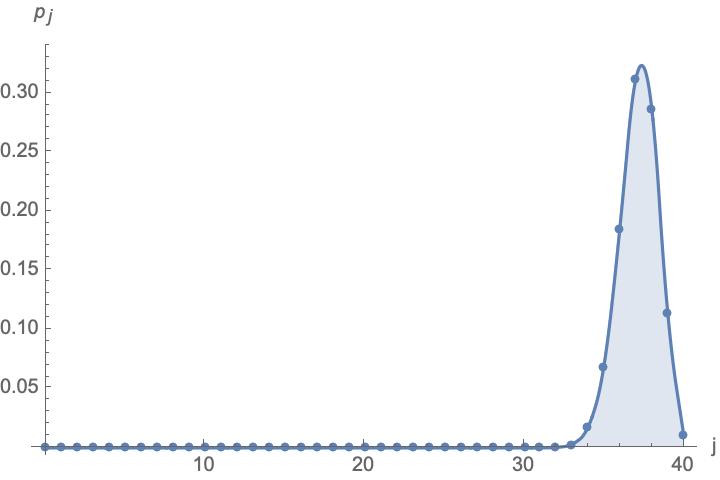}}  
    \\    
    \subfloat[$\vec{\theta}=(\frac{\pi}{2},\frac{\pi}{2},\frac{\pi}{6},\frac{5\pi}{6}),  \vec{\varphi}=(0,\frac{2\pi}{3},\frac{\pi}{3},\frac{\pi}{3})$]{\label{2d3}\includegraphics[width=0.5\linewidth]{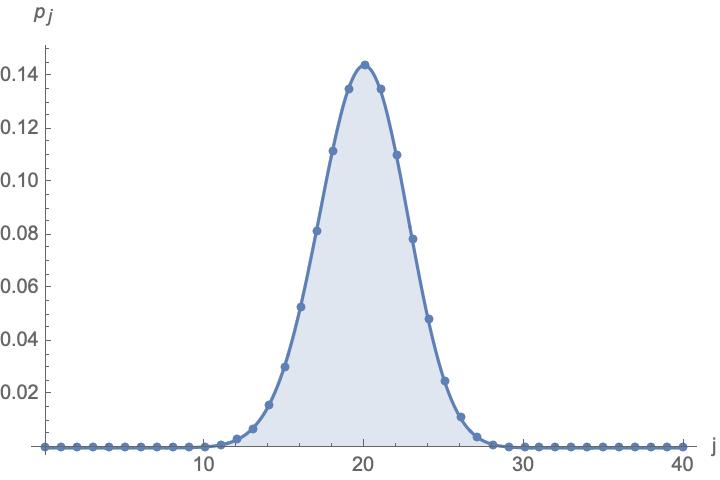}}  
    \subfloat[$\vec{\theta}=(\frac{\pi}{2},\frac{\pi}{2},\frac{\pi}{16},\frac{15\pi}{16}),  \vec{\varphi}=(0,\frac{7\pi}{8},\frac{7\pi}{16},\frac{7\pi}{16})$]{\label{7d8}\includegraphics[width=0.5\linewidth]{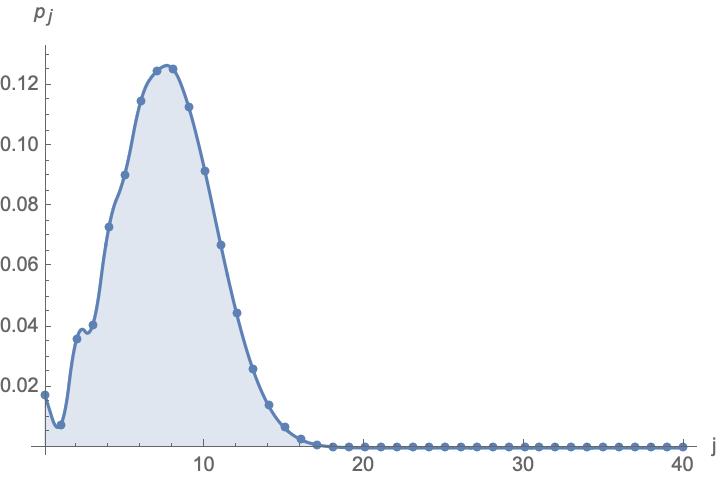}}   
     \caption{The numerical results of $p_j$ for $j_1=j_2=j_3=j_4=20$ and various $\vec{\theta}\equiv(\theta_1,\theta_2,\theta_3,\theta_4), \vec{\varphi}\equiv(\varphi_1,\varphi_2,\varphi_3,\varphi_4)$, where the $x$-axis shows the recoupling spin $j$ and $y$-axis shows the numerical value of $p_j$. These figures show that $p_j$ has a peak near $j=j_0\equiv|j_1\hat{n}_1+j_2\hat{n}_2|$, and this peak shrinks with the angle  $\arccos{(\hat{n}_1\cdot\hat{n}_2)}$ decreasing.}
     \label{fig2}    
\end{figure}

One can see that $E(A|B)$ is determined by the distribution $p_j$. Generally, $p_j$ is a rather complicated function of $j$ for given $(j_1,\hat{n}_1,j_2,\hat{n}_2,j_3,\hat{n}_3,j_4,\hat{n}_4)$, thus it is hard to analyze the property of $p_j$ by its analytical expression.  We calculate  $p_j$ and $E(A|B)$ by the numerical methods as shown in Tab.\ref{tab1}, Tab.\ref{tab2}, Fig.\ref{fig1} and Fig.\ref{fig2}. Our results shows that, $p_j$ has a peak near $j=j_0\equiv|j_1\hat{n}_1+j_2\hat{n}_2|$. Especially, this peak shrinks relative to the  range of $j$ with the boundary spins $j_1,j_2,j_3,j_4$ getting larger,  and this peak shrinks with the angle $\arccos(\hat{n}_1\cdot \hat{n}_2)$ decreasing. Thus, we can argue that $E_0(A|B)=\sum_{j}p_j S^A_j\approx\ln(2j_0+1)$ reasonably. The term $E_p(A|B)$ is just the Shannon entropy of the distribution $p_j$. Note that the range $\max(|j_1-j_2|,|j_3-j_4|)\leq j\leq \min(|j_1+j_2|,|j_3+j_4|)$ of $j$ grows linearly with $j_1,j_2,j_3,j_4$ going large. Then, based on the peakedness property of $p_j$, one can conclude that $E_p(A|B)$ increases no more than logarithmic growth with $j_1,j_2,j_3,j_4$ going large.



\begin{table}
    \centering
    \begin{tabular}{|c|c|c|c|l|l|} \hline  
         $(j_1,j_2,j_3,j_4)$&  $(\theta_1,\theta_2,\theta_3,\theta_4)$&  $(\varphi_1,\varphi_2,\varphi_3,\varphi_4)$&  $E_p(A|B)$&$E_0(A|B)$&$E(A|B)$\\ \hline  
         $(5,5,5,5)$&  $(\frac{\pi}{2},\frac{\pi}{2},\frac{\pi}{4},\frac{3\pi}{4})$&  $(0,\frac{\pi}{2},\frac{\pi}{4},\frac{\pi}{4})$&  1.56229&2.70936&4.27165\\ \hline  
         $(10,10,10,10)$&  $(\frac{\pi}{2},\frac{\pi}{2},\frac{\pi}{4},\frac{3\pi}{4})$&  $(0,\frac{\pi}{2},\frac{\pi}{4},\frac{\pi}{4})$&  1.8922&3.37298&5.26517\\ \hline  
         $(15,15,15,15)$& $(\frac{\pi}{2},\frac{\pi}{2},\frac{\pi}{4},\frac{3\pi}{4})$&  $(0,\frac{\pi}{2},\frac{\pi}{4},\frac{\pi}{4})$&  2.08976&3.76835&5.85811\\ \hline  
         $(20,20,20,20)$&  $(\frac{\pi}{2},\frac{\pi}{2},\frac{\pi}{4},\frac{3\pi}{4})$&   $(0,\frac{\pi}{2},\frac{\pi}{4},\frac{\pi}{4})$&  2.23109&4.05092&6.28201\\ \hline 
 $(25,25,25,25)$& $(\frac{\pi}{2},\frac{\pi}{2},\frac{\pi}{4},\frac{3\pi}{4})$&  $(0,\frac{\pi}{2},\frac{\pi}{4},\frac{\pi}{4})$& 2.34114&4.271&6.61214\\\hline 
 $(30,30,30,30)$& $(\frac{\pi}{2},\frac{\pi}{2},\frac{\pi}{4},\frac{3\pi}{4})$&  $(0,\frac{\pi}{2},\frac{\pi}{4},\frac{\pi}{4})$& 2.43131&4.45127&6.88258\\ \hline
    \end{tabular}
    \caption{The numerical values of $E_p(A|B)$ and $E_0(A|B)$ for fixed $\hat{n}_1, \hat{n}_2,\hat{n}_3,\hat{n}_4$ and growth boundary spins. It is shown that $E_p(A|B)$ and $E_0(A|B)$ both grow with the boundary spins $j_1=j_2=j_3=j_4$ getting larger.}
    \label{tab1}
\end{table}

\begin{table}
    \centering
    \begin{tabular}{|c|c|c|c|l|l|} \hline  
         $(j_1,j_2,j_3,j_4)$&  $(\theta_1,\theta_2,\theta_3,\theta_4)$&  $(\varphi_1,\varphi_2,\varphi_3,\varphi_4)$&  $E_p(A|B)$&$E_0(A|B)$&$E(A|B)$\\ \hline  
 $(20,20,20,20)$& $(\frac{\pi}{2},\frac{\pi}{2},\frac{7\pi}{16},\frac{9\pi}{16})$& $(0,\frac{\pi}{8},\frac{\pi}{16},\frac{\pi}{16})$& 0.861644&4.38243&5.24407\\ \hline 
         $(20,20,20,20)$&  $(\frac{\pi}{2},\frac{\pi}{2},\frac{3\pi}{8},\frac{5\pi}{8})$&  $(0,\frac{\pi}{4},\frac{\pi}{8},\frac{\pi}{8})$&  1.6089&4.3212&5.9301\\ \hline  
         $(20,20,20,20)$&  $(\frac{\pi}{2},\frac{\pi}{2},\frac{\pi}{4},\frac{3\pi}{4})$&  $(0,\frac{\pi}{2},\frac{\pi}{4},\frac{\pi}{4})$&  2.23109&4.05092&6.28201\\ \hline 
 $(20,20,20,20)$& $(\frac{\pi}{2},\frac{\pi}{2},\frac{\pi}{6},\frac{5\pi}{6})$& $(0,\frac{2\pi}{3},\frac{\pi}{3},\frac{\pi}{3})$& 2.43579&3.69809&6.13388\\\hline  
 $(20,20,20,20)$& $(\frac{\pi}{2},\frac{\pi}{2},\frac{\pi}{8},\frac{7\pi}{8})$&$ (0,\frac{3\pi}{4},\frac{3\pi}{8},\frac{3\pi}{8}) $& 2.50054&3.42048&5.92102\\ \hline 
 $(20,20,20,20)$& $(\frac{\pi}{2},\frac{\pi}{2},\frac{\pi}{16},\frac{15\pi}{16})$& $(0,\frac{7\pi}{8},\frac{7\pi}{16},\frac{7\pi}{16})$& 2.541&2.65329&5.1943\\ \hline 
    \end{tabular}
    \caption{The numerical values of $E_p(A|B)$ and $E_0(A|B)$ for fixed boundary spins $j_1=j_2=j_3=j_4=20$ and various angle $\arccos{(\hat{n}_1\cdot\hat{n}_2)}$. It is shown that $E_p(A|B)$ increases while $E_0(A|B)$ decreases with the angle $\arccos{(\hat{n}_1\cdot\hat{n}_2)}$ getting larger.}
    \label{tab2}
\end{table}

\subsubsection{Coherent intertwiner with arbitrary number of legs}\label{sec3.3.2}
Let us consider the gauge invariant coherent intertwiner space $\mathcal{H}_v^{\{j,\tilde{j}\}}=\text{Inv}_{SU(2)}[\bigotimes_{I=1}^{P}V^{j_I} \otimes\bigotimes_{J=1}^{Q}\bar{V}^{\tilde{j}_J}]$ on a $(P+Q)$-valent vertex, in which an element is given by
\begin{equation}\label{Ici}
|\mathcal{I}\rangle=\int_{SU(2)}dg \bigotimes_{I=1}^{P}g|j_I,\hat{n}_I\rangle\bigotimes_{J=1}^{Q}\langle \tilde{j}_J,\hat{\tilde{n}}_J|g^{-1}.
\end{equation}
 Similar to the coherent intertwiner with 4-legs, the coherent intertwiner with arbitrary number of legs can be expanded by the orthogonal re-coupling basis of intertwiner space, which reads
\begin{eqnarray}
&&|\mathcal{I}\rangle\\\nonumber
   &=&\sum_{j}(2j+1)\Bigg(\sum_{k,m=-j}^j |\mathcal{I}_{P};j,m,k\rangle\langle\mathcal{I}_{Q};j,m,k|\Bigg)\\\nonumber
   &=&\sum_{j}\sum_{j_{\imath_1}}\sum_{j_{\imath_2}}...\sum_{j_{\imath_{p}}} \sum_{\tilde{j}_{\imath_1}}\sum_{\tilde{j}_{\imath_2}}...\sum_{\tilde{j}_{\imath_{q}}} (2j+1)^{-1} \cdot c^{\mathcal{I}_{P},\mathcal{I}_{Q}}_{j,j_\imath,\tilde{j}_\imath}\\\nonumber
   &&\cdot\sum_{k} \sum_{k_{\imath_1},...,k_{\imath_p}}\sum_{\tilde{k}_{\imath_1},...,\tilde{k}_{\imath_q}}\Bigg( |j_{P},j_{P-1};j_{\imath_p},k_{\imath_p}\rangle \otimes|j_{\imath_p},k_{\imath_p};j_p;j_{\imath_{p-1}},k_{\imath_{p-1}}\rangle\otimes...  \\\nonumber
   &&\otimes|j_{\imath_2},k_{\imath_2};j_2;j_{\imath_1},k_{\imath_1}\rangle \otimes|j_{\imath_1},k_{\imath_1};j_1;j,k\rangle \langle\tilde{j}_{Q},\tilde{j}_{Q-1};\tilde{j}_{\imath_q},\tilde{k}_{\imath_q}| \otimes\langle \tilde{j}_{\imath_q},\tilde{k}_{\imath_q};\tilde{j}_q;\tilde{j}_{\imath_{q-1}},\tilde{k}_{\imath_{q-1}}| \otimes... \\\nonumber
   &&\otimes\langle \tilde{j}_{\imath_2},\tilde{k}_{\imath_2};\tilde{j}_2;\tilde{j}_{\imath_1},\tilde{k}_{\imath_1}| \otimes\langle \tilde{j}_{\imath_1}, \tilde{k}_{\imath_1};\tilde{j}_1;j,k|\Bigg),
\end{eqnarray}
where  $|j',j'';j,k\rangle:=\sum_{m',m''}C^{j'j''j}_{m'm''k}|j',m'\rangle \otimes|j'',m''\rangle$,  $|j',k';j'';j,k\rangle:=\sum_{m''}C^{j'j''j}_{k'm''k}|j'',m''\rangle$,
and $c^{\mathcal{I}_{P},\mathcal{I}_{Q}}_{j,j_\imath,\tilde{j}_\imath}$ is the coefficient of this expansion, see the details in Appendix \ref{app2}.

Now, the density matrix of the coherent intertwiner $|\mathcal{I}\rangle$ can be given by $\rho_{\mathcal{I}}\equiv\frac{|\mathcal{I}\rangle\langle \mathcal{I}|}{\langle \mathcal{I}|\mathcal{I}\rangle}$.
Let us define $\mathcal{H}_A:=\otimes_{I=1}^{P}V^{j_I} $ and $\mathcal{H}_B:= \otimes_{J=1}^{Q}\bar{V}^{\tilde{j}_J}$. Then,  the entanglement $E(A|B)$ between $A$ and $B$ can be given by the Von Neumann entropy of the reduced density matrices $\rho_{A}$, which are defined as
\begin{equation}
\rho_{A}:=\text{Tr}_B(\rho_{\mathcal{I}}),
\quad E(A|B):=-\text{Tr}_A(\rho_{A}\ln \rho_{A}).
\end{equation}
More explicitly, $\rho_{A}$ can be calculated as
\begin{eqnarray}
\rho_{A}&=&\text{Tr}_B(\rho_{\mathcal{I}})=\frac{\text{Tr}_B(|\mathcal{I}\rangle\langle \mathcal{I}|)}{\langle \mathcal{I}|\mathcal{I}\rangle},
\end{eqnarray}
where
\begin{eqnarray}
&&\langle \mathcal{I}|\mathcal{I}\rangle   \\\nonumber
  &=&\sum_{j}\sum_{j_{\imath_1},j_{\imath_2},...,j_{\imath_{p}}} \sum_{\tilde{j}_{\imath_1},\tilde{j}_{\imath_1},...,\tilde{j}_{\imath_q}} (2j+1) ^{-1}c^{\mathcal{I}_{P},\mathcal{I}_{Q}}_{j,j_\imath,\tilde{j}_\imath} \overline{c^{\mathcal{I}_{P},\mathcal{I}_{Q}}_{j,j_\imath,\tilde{j}_\imath}}.
\end{eqnarray}
and
\begin{eqnarray}
&&\text{Tr}_B(|\mathcal{I}\rangle\langle \mathcal{I}|)   \\\nonumber
   &=&\sum_{j}\sum_{j_{\imath_1},j_{\imath_2},...,j_{\imath_{p}}} \sum_{j'_{\imath_1},j'_{\imath_2},...,j'_{\imath_{p}}}  (2j+1) ^{-1}\sum_{\tilde{j}_{\imath_1},\tilde{j}_{\imath_1},...,\tilde{j}_{\imath_q}}c^{\mathcal{I}_{P},\mathcal{I}_{Q}}_{j,j_\imath,\tilde{j}_\imath} \overline{c^{\mathcal{I}_{P},\mathcal{I}_{Q}}_{j,j_\imath,\tilde{j}_\imath}}\rho_{j,j_{\imath},\tilde{j}_{\imath}}^{A},
\end{eqnarray}
with 
 \begin{eqnarray}
\rho_{j,j_{\imath},\tilde{j}_{\imath}}^{A}&:=&  (2j+1) ^{-1} \cdot\sum_{k} \sum_{k_{\imath_1},...,k_{\imath_p}} \sum_{k'_{\imath_1},...,k'_{\imath_p}}\Bigg( |j_{P},j_{P-1};j_{\imath_p},k_{\imath_p}\rangle \otimes|j_{\imath_p},k_{\imath_p};j_p;j_{\imath_{p-1}},k_{\imath_{p-1}}\rangle\otimes...  \\\nonumber
   &&\otimes|j_{\imath_2},k_{\imath_2};j_2;j_{\imath_1},k_{\imath_1}\rangle \otimes|j_{\imath_1},k_{\imath_1};j_1;j,k\rangle \langle{j}_{P},{j}_{P-1};j'_{\imath_p},k'_{\imath_p}| \otimes\langle j'_{\imath_p},k'_{\imath_p};{j}_p;j'_{\imath_{p-1}},k'_{\imath_{p-1}}| \otimes... \\\nonumber
   &&\otimes\langle j'_{\imath_2},k'_{\imath_2};j_2;j'_{\imath_1},k'_{\imath_1}| \otimes\langle j'_{\imath_1}, k'_{\imath_1};j_1;j,k|\Bigg).
\end{eqnarray}
 It is direct to calculate that the Von Neumann entropy of $\rho_{j,j_{\imath},\tilde{j}_{\imath}}^{A}$, which leads
 \begin{equation}
 S^A_{j}:=-\text{Tr}_A(\rho_{j,j_{\imath},\tilde{j}_{\imath}}^{A}\ln \rho_{j,j_{\imath},\tilde{j}_{\imath}}^{A})=\ln(2j+1).
 \end{equation}
 Further, let us define
 \begin{eqnarray}
p_{j,j_{\imath},j'_{\imath}}  &:=& \frac{ (2j+1) ^{-1}\sum_{\tilde{j}_{\imath_1},\tilde{j}_{\imath_1},...,\tilde{j}_{\imath_q}}c^{\mathcal{I}_{P},\mathcal{I}_{Q}}_{j,j_\imath,\tilde{j}_\imath} \overline{c^{\mathcal{I}_{P},\mathcal{I}_{Q}}_{j,j'_\imath,\tilde{j}_\imath}}}{\langle \mathcal{I}|\mathcal{I}\rangle  },
 \end{eqnarray}
 \begin{equation}
\tilde{p}_j:=\sum_{j_{\imath_1},j_{\imath_2},...,j_{\imath_{p}}} \sum_{j'_{\imath_1},j'_{\imath_2},...,j'_{\imath_{p}}}  p_{j,j_{\imath},j'_{\imath}},
\end{equation}
and
\begin{equation}
\bar{p}_{j,j_{\imath},j'_{\imath}} :=\frac{p_{j,j_{\imath},j'_{\imath}} }{\tilde{p}_j}.
\end{equation}
Then, the entanglement entropy $E(A|B)$ can be given by
\begin{equation}
E(A|B)=-\text{Tr}_A(\rho_{A}\ln \rho_{A})=E_p(A|B)+\sum_{j}\tilde{p}_j \ln(2j+1)
\end{equation}
where 
\begin{eqnarray}
E_p(A|B)&:=&-\sum_{j}\sum_{j_{\imath_1},j_{\imath_2},...,j_{\imath_{p}}} \sum_{j'_{\imath_1},j'_{\imath_2},...,j'_{\imath_{p}}}(p_{j,j_{\imath},j'_{\imath}}\ln p_{j,j_{\imath},j'_{\imath}})\\\nonumber
&=&-\sum_{j}\tilde{p}_j\ln \tilde{p}_j+\sum_{j}\tilde{p}_jE_{\bar{p}_j}(A|B)
\end{eqnarray}
 with
\begin{equation}
E_{\bar{p}_j}(A|B):=-\sum_{j_{\imath_1},j_{\imath_2},...,j_{\imath_{p}}} \sum_{j'_{\imath_1},j'_{\imath_2},...,j'_{\imath_{p}}} (\bar{p}_{j,j_{\imath},j'_{\imath}}\ln \bar{p}_{j,j_{\imath},j'_{\imath}}).
\end{equation}

It is worth to have a discussion on this result. First, one can notice that $E(A|B)$ contains three terms $\sum_{j}\tilde{p}_jE_{\bar{p}_j}(A|B)$, $\sum_{j}\tilde{p}_j \ln(2j+1)$ and $-\sum_{j}\tilde{p}_j\ln \tilde{p}_j$. This result takes a same formulation as the result \eqref{3terms} for the boundary entanglement of two entangled intertwiners. In fact, the recoupling edge labelled by spin $j$ separates the coherent intertwiner as two entangled intertwiners, and the corresponding boundary entanglement of these two entangled intertwiners is given by $E(A|B)$ exactly, with $E_{p}(A|B)$ being the intertwiner entanglement and $\tilde{p}_j$ being the probability distribution of the spin $j$ on recoupling edge. 
 Second, it is easy to see that the entanglement entropy $E(A|B)$ depends on the distribution $p_{j,j_{\imath},\tilde{j}_{\imath}}=\tilde{p}_j\cdot\bar{p}_{j,j_{\imath},\tilde{j}_{\imath}}$ on the spins of the recoupling edges. In fact, by considering the peakedness property of coherent intertwiner, one can argue that  $p_{j,j_{\imath},\tilde{j}_{\imath}}$ is peaked near the values
\begin{eqnarray}
j_{\imath_p}=|j_{P}\hat{n}_{P}+j_{{P-1}}\hat{n}_{{P-1}}|,&& j_{\imath_{p-1}}=|j_{P}\hat{n}_{P}+j_{{P-1}}\hat{n}_{{P-1}}+j_p\hat{n}_p|,\\\nonumber
...,\\\nonumber
j_{\imath_1}=|j_{P}\hat{n}_{P}+j_{{P-1}}\hat{n}_{{P-1}}+j_p\hat{n}_p+...+j_2\hat{n}_2|,&& j=|j_{P}\hat{n}_{P}+j_{{P-1}}\hat{n}_{{P-1}}+j_p\hat{n}_p+...+j_1\hat{n}_1|,\\\nonumber
\tilde{j}_{\imath_q}=|\tilde{j}_{Q}\hat{\tilde{n}}_{Q}+\tilde{j}_{{Q-1}}\hat{\tilde{n}}_{{Q-1}}|,&& \tilde{j}_{\imath_{q-1}}=|\tilde{j}_{Q}\hat{\tilde{n}}_{Q}+\tilde{j}_{{Q-1}}\hat{\tilde{n}}_{{Q-1}} +\tilde{j}_q\hat{\tilde{n}}_q|,\\\nonumber
...,\\\nonumber
\tilde{j}_{\imath_1}=|\tilde{j}_{Q}\hat{\tilde{n}}_{Q}+\tilde{j}_{{Q-1}} \hat{\tilde{n}}_{{Q-1}}+\tilde{j}_q\hat{\tilde{n}}_q+...+\tilde{j}_2\hat{\tilde{n}}_2|,&& j=|\tilde{j}_{Q}\hat{\tilde{n}}_{Q}+\tilde{j}_{{Q-1}}\hat{\tilde{n}}_{{Q-1}} +\tilde{j}_q\hat{\tilde{n}}_q+...+\tilde{j}_1\hat{\tilde{n}}_1|.
\end{eqnarray}
This argument could be checked by evaluating the specific property of  the distribution $p_{j,j_{\imath},\tilde{j}_{\imath}}$. However, one can see that the general expression of  $p_{j,j_{\imath},\tilde{j}_{\imath}}$ is too complicated to proceed the analytical study. One may also expect a numerical calculation of the distribution $p_{j,j_{\imath},\tilde{j}_{\imath}}$ , and we would like to leave this to future researches.


\section{Conclusion and Outlook}\label{sec4}

To summarize, we focus on the entanglement 
between the legs of the intertwiners on  vertices.  We first review the relation between  the boundary entanglement and intertwiner entanglement, and extend the result to the  case that the internal edge carries a spin-superposition.  Then,  we turn to consider the specific intertwiners on a four-valents vertex, which is decomposed as two parts $A$ and $B$ attached by the labels $(j_1m_1,j_2m_2)$ and  $(j_3m_3,j_4m_4)$ respectively.  By introducing the group-averaging to the tensor-product type intertwiner on  the four-valents vertex,  we calculate the entanglement entropy $E(A|B)$ encoded in the group-averaged tensor-product intertwiner with various weights. The results show that, for the group-averaged tensor-product intertwiner with highest (and lowest) weight,  the entanglement entropy is able to capture the main character of the probability distribution $p_j$ of the recoupling spin $j$ . This result suggests us  to consider the entanglement entropy encoded in the gauge invariant coherent intertwiners. 
By introducing the recoupling edges to decompose the coherent intertwiner labelled by $(j_1\hat{n}_1,j_2\hat{n}_2)$ and  $(j_3\hat{n}_3,j_4\hat{n}_4)$, we find that the entanglement is determined by the probability distribution $p_{j}$ of the spins $j$ on the recoupling edges of the coherent intertwiner, and the result of entanglement $E(A|B)$ is composed by the sum of two terms $E_p(A|B)$ and $E_0(A|B)$. The first term $E_p(A|B)$ is just the Shannon entropy
of the distribution $p_{j}$, while the second term $E_0(A|B)$ is the expectation value of $\ln(2j+1)$ with respect to the distribution $p_{j}$.    Our results show that, $p_j$ has a peak near $j=j_0\equiv|j_1\hat{n}_1+j_2\hat{n}_2|$; Especially, this peak shrinks relative to the  range of $j$ with the boundary spins $j_1,j_2,j_3,j_4$ getting larger,  and this peak shrinks with the angle $\arccos(\hat{n}_1\cdot \hat{n}_2)$ decreasing.
Thus, we  argue that $E_0(A|B)=\sum_{j}p_j S^A_j\approx\ln(2j_0+1)$, and the term $E_p(A|B)$ increases no more than logarithmic growth with $j_1,j_2,j_3,j_4$ going large.
We also extend the analytical calculation part of  entanglement  to the case of
 gauge invariant coherent intertwiner on a $(P+Q)$-valents vertex, which is also separated as two parts $A$ and $B$ attached by $(j_1,\hat{n}_1,j_2,\hat{n}_2,...,j_P,\hat{n}_P)$ and  $(\tilde{j}_1,\hat{\tilde{n}}_1,\tilde{j}_2,\hat{\tilde{n}}_2,...,\tilde{j}_Q,\hat{\tilde{n}}_Q)$ respectively. By introducing the recoupling edges to decompose the  gauge invariant  coherent intertwiner, we give the probability distribution $p_{j,j_\imath,\tilde{j}_\imath}$ of the spins $j,j_\imath,\tilde{j}_\imath$ on the recoupling edges   analytically, and  show that the entanglement between $A$ and $B$ is determined by $p_{j,j_\imath,\tilde{j}_\imath}$.

It is worth to have some discussion on the results. First, one should notice that the entanglement entropy $E(A|B)$ are given based on the gauge invariant coherent intertwiners, which belong to the gauge invariant subspace of the total system $\mathcal{H}_A\otimes \mathcal{H}_B$. Thus, $E(A|B)$ describes the entanglement between some gauge invariant degrees of freedom, e.g. the face-angle $\theta=\arccos{\hat{n}_1\cdot \hat{n}_2}$. However, the appearance of the factor $\ln(2j+1)$ comes from breaking the gauge invariance, and the physical meaning of the factor  $\ln(2j+1)$ depends on how to define boundaries at the quantum level for non-Abelian lattice gauge theories and LQG \cite{Livine:2017fgq}. Second, the entanglement  between the legs of coherent intertwiner can be related to the face-angle of the semiclassical polyhedron. Recall that $p_j$ has a peak near $j_0\equiv|j_1\hat{n}_1+j_2\hat{n}_2|$  and this peak shrinks with the angle $\arccos(\hat{n}_1\cdot \hat{n}_2)$ decreasing. Thus, the Shannon entropy  $E_p(A|B)$ 
of the distribution $p_{j}$  decreases with the face-angle $\pi-\arccos(\hat{n}_1\cdot \hat{n}_2)$ increasing. Nevertheless, one also note that another term $E_0(A|B)\approx\ln(2j_0+1)$ in $E(A|B)$ increases with the face-angle $\pi-\arccos(\hat{n}_1\cdot \hat{n}_2)$ increasing. Hence, the entanglement $E(A|B)$ may not have monotonous dependency on the face-angle $\pi-\arccos(\hat{n}_1\cdot \hat{n}_2)$, e.g. as shown in Tab.\ref{tab2}, $E(A|B)$ increases first and then decreases with the face-angle $\pi-\arccos(\hat{n}_1\cdot \hat{n}_2)$ increasing for $j_1=j_2=j_3=j_4=20$. 
Third, we would like to emphasis that the entanglement on the two-vertices graph with one link is  independent on holonomy-insertion \cite{Livine:2017fgq}. In other words, the holonomy living along the edge connecting these two vertices is irrelevant to the entanglement entropy at all. This property is attributed to the observation that the holonomy can be eliminated by a boundary unitary. This implies that the entanglement do not distinguish between a two-vertex graph and its coarse-grained graph (single vertex) provided that there is only one link.
For the cases of two-vertices graph with multi-links, the nontrivial loop is introduced, which should be viewed as excitation of gauge curvature \cite{Freidel:2002xb,Charles:2016xwc}, or interpreted as topological defect \cite{Deser:1983tn,Freidel:2005me,Livine:2008iq}. The related studies of entanglement for the cases can be found in \cite{Livine:2008iq,Chen:2022rty}. As for the entanglement in coherent states admitting gauge curvature, we leave the exploration in the future.

It is also worth to have an outlook on the
use of entanglement  in  the study of black hole entropy in the framework of loop
quantum gravity \cite{Ashtekar:1997yu,Ashtekar:2000eq,Engle:2009vc,Song:2020arr,Ghosh:2011fc,Song:2022zit}. A typical method for statistics of black holes in loop quantum gravity is to
consider the classical boundary conditions for isolated horizons \cite{Ashtekar:1997yu,Engle:2009vc}. This boundary theory is given as
a Chern-Simons theory and the horizon is described as a classical surface punctured by the edges of spin-network states. Then,  the black hole entropy is calculated by counting the Chern-Simons states for a punctured sphere. This picture could also be considered in the full theory of LQG.  We would have the horizon cutting the spin-networks, and the entanglement entropy between the black hole interior and exterior comes directly from the contributions of all the punctures on the
horizon \cite{Donnelly:2008vx,Perez:2014ura}. In fact, these punctures can be generated by the edges or true vertices of the spin-networks crossing the horizon. It has been argued that the puncture generated by an edge crossing the horizon contributes the boundary spin state entanglement to the entanglement entropy between the black hole interior and
exterior \cite{Livine:2017fgq}. Nevertheless, the space-time geometry near the horizon of a massive Schwarzschild black hole should be  described by the semi-classical state in LQG, which could not only control the classical property of horizon but also admit the quantum perturbation to generate the Hawking radiation. Thus,  it seems inevitable that the coherent states on graphs and coherent intertwiners on vertices would control the entanglement contribution of the punctures generated by the edges or true vertices crossing the horizon. Our studies in this paper provides a basement to explore the entanglement contribution coming from coherent intertwiners, and it can be extended to study the the entanglement contribution coming from coherent states on graphs (e.g., Thiemann's coherent state and twisted geometry coherent state \cite{Thomas2001Gauge,2001Gauge,Bianchi:2009ky,Calcinari_2020,PhysRevD.104.046014,1994The,Long:2021lmd,Long:2022cex}).




\section*{Acknowledgments}
This work is supported by the project funded by  the National Natural Science Foundation of China (NSFC) with Grants No. 12047519 and No. 12165005. G. L. is supported by ``the Fundamental Research Funds for the Central Universities''. Q. C. is funded within the QuantERA II Programme that has received funding from the European Union’s Horizon 2020 research and innovation programme under Grant Agreement No 101017733 (VERIqTAS).

\begin{appendix}





\section{The re-coupling expansion of coherent intertwiner with arbitrary number of legs} \label{app2}

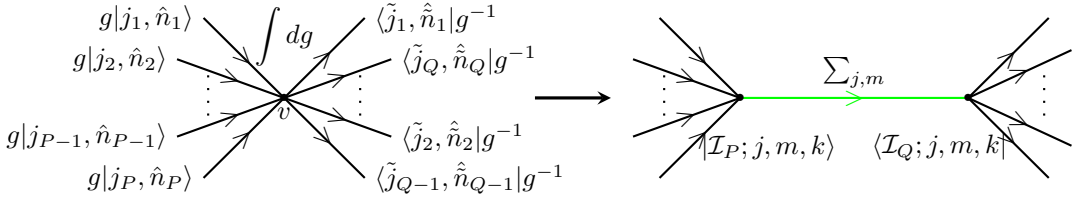
\begin{figure}[htb]
	\centering
	\begin{tikzpicture} [scale=1]
\coordinate  (O) at (0,0);

\coordinate  (O1) at (-1,0.3);

\coordinate  (O2) at (-1,-0.3);

\coordinate  (O3) at (1,0.3);

\coordinate  (O4) at (1,-0.3);

\coordinate  (A) at (6,0);

\coordinate  (A1) at (5,0.3);

\coordinate  (A2) at (5,-0.3);

\coordinate  (B) at (9,0);

\coordinate  (B1) at (10,0.3);

\coordinate  (B2) at (10,-0.3);

\node[scale=0.7] at (O) {$\bullet$} node[below] {$v$}node [above=0.3] {$\displaystyle{ \int d g }$};

\draw[thick] (O)  --  node[midway,sloped]{$>$} ++ (45:1.5) node[right] {$\langle \tilde{j}_1, \hat{\tilde{n}}_1 | g^{-1} $};

\draw[thick] (O)  to  node[midway,sloped]{$>$} ++ (20:1.5) node[right] {$\langle \tilde{j}_{Q}, \hat{\tilde{n}}_{Q} | g^{-1} $};

\draw[thick] (O)  --  node[midway,sloped]{$>$} ++ (-20:1.5) node[right] {$\langle \tilde{j}_2, \hat{\tilde{n}}_2 | g^{-1} $};

\draw[thick] (O)  to  node[midway,sloped]{$>$} ++ (-45:1.5) node[right] {$\langle \tilde{j}_{Q-1}, \hat{\tilde{n}}_{Q-1} | g^{-1} $};

\draw[thick] (O)  to  node[midway,sloped]{$>$} ++ (200:1.5) node[left] {$g | j_{P-1}, \hat{n}_{P-1} \rangle$};

\draw[thick] (O)  to  node[midway,sloped]{$>$} ++ (160:1.5) node[left]{$g | j_2, \hat{n}_2 \rangle$} ;
\draw[thick] (O)  to  node[midway,sloped]{$>$} ++ (225:1.5) node[left] {$g | j_P, \hat{n}_P \rangle$} ;

\draw[thick] (O)  to  node[midway,sloped]{$>$} ++ (135:1.5) node[left]{$g | j_1, \hat{n}_1 \rangle$};

\draw [thick, loosely dotted] (O1) -- (O2);
\draw [thick, loosely dotted] (O3) -- (O4);

\draw [thick, loosely dotted] (A1) -- (A2);
\draw [thick, loosely dotted] (B1) -- (B2);

\draw[->,>=stealth,very thick] (O) ++ (3.3,0) -- ++ (1,0);

\draw[thick,green] (A) -- node[midway,sloped]{$>$} node[above,black] {$\sum_{j,m}$} (B);

\draw[thick] (B) ++ (240:0.75) node [black] {$\langle\mathcal{I}_Q;j,m,k|$};
\draw[thick] (A) ++ (300:0.75) node [black] {$|\mathcal{I}_P;j,m,k \rangle$};

\draw[thick] (B)  --  node[midway,sloped]{$>$} ++ (45:1.5)  ;

\draw[thick] (B)  to  node[midway,sloped]{$>$} ++ (25:1.5);
\draw[thick] (B)  to  node[midway,sloped]{$>$} ++ (-25:1.5) ;

\draw[thick] (B)  to  node[midway,sloped]{$>$} ++ (-45:1.5);
\draw[thick] (A)  to  node[midway,sloped]{$>$} ++ (135:1.5) ;

\draw[thick] (A)  to  node[midway,sloped]{$>$} ++ (160:1.5);

\draw[thick] (A)  to  node[midway,sloped]{$>$} ++ (200:1.5);

\draw[thick] (A)  to  node[midway,sloped]{$>$} ++ (225:1.5);

\node[scale=0.7] at (A) {$\bullet$};
\node[scale=0.7] at (B) {$\bullet$};

\end{tikzpicture}
\caption{The illustration of recoupling spin $j$ for $\mathcal{I}$.}
\label{fig:AnyValency-vertex}
\end{figure}

By introduce a recoupling edge labelled by spin $j$, the coherent intertwiner $|\mathcal{I}\rangle$ can be decomposed as
\begin{eqnarray}
&&|\mathcal{I}\rangle\\
\nonumber
&=&\int_{SU(2)}dg\bigotimes_{I=1}^{P}g|j_I,\hat{n}_I\rangle\int_{SU(2)}dh\delta(g^{-1}h)\bigotimes_{J=1}^{Q}\langle \tilde{j}_J,\hat{\tilde{n}}_J|h^{-1} \\\nonumber
   &=&\sum_{j}(2j+1) \int_{SU(2)}dg\int_{SU(2)}dh\bigotimes_{I=1}^{P}g|j_I,\hat{n}_I\rangle\text{tr}^{j}(g^{-1}h)\bigotimes_{J=1}^{Q}\langle\tilde{ j}_J,\hat{\tilde{n}}_J|h^{-1} \\\nonumber
   &=&\sum_{j}(2j+1)\Bigg(\sum_{m,k=-j}^j \int_{SU(2)}dg\bigotimes_{I=1}^{P}g|j_I,\hat{n}_I\rangle\otimes \langle j,m|g^{-1}|j,k\rangle\otimes\int_{SU(2)}dh\langle j,k| h| j,m\rangle\bigotimes_{J=1}^{Q}\langle\tilde{ j}_J,\hat{\tilde{n}}_J|h^{-1} \Bigg)
   \\\nonumber
   &=&\sum_{j}(2j+1)\Bigg(\sum_{m,k=-j}^j |\mathcal{I}_{P};j,m,k\rangle\langle\mathcal{I}_{Q};j,m,k|\Bigg),
\end{eqnarray}
where
\begin{eqnarray}
&&|\mathcal{I}_{P};j,m,k\rangle:=\int_{SU(2)}dg\bigotimes_{I=1}^{P}g|j_I,\hat{n}_I\rangle\otimes \langle j,m|g^{-1}|j,k\rangle,
\\\nonumber
&&
\langle\mathcal{I}_{Q};j,m,k|:=\int_{SU(2)}dh \langle j,k|h| j,m\rangle\bigotimes_{J=1}^{Q}\langle \tilde{j}_J,\hat{\tilde{n}}_J|h^{-1}
\end{eqnarray}
are projectors  that send tensor representations to recoupling representations, see the illustration in Fig.\ref{fig:AnyValency-vertex}.

%
\begin{figure}[htb]
	\centering
	\begin{tikzpicture} [scale=1]

\coordinate  (O) at (0,0);

\coordinate  (O1) at (-1,0.3);

\coordinate  (O2) at (-1,-0.3);

\draw [thick, loosely dotted] (O1) -- (O2);

\node[scale=0.7] at (O) {$\bullet$} ++ (300:0.75) node [black] {$|\mathcal{I}_P;j,m,k\rangle$};

\draw[thick] (O)  to  node[midway,sloped]{$>$} ++ (135:1.5);

\draw[thick] (O)  to  node[midway,sloped]{$>$} ++ (160:1.5);

\draw[thick] (O)  to  node[midway,sloped]{$>$} ++ (200:1.5);

\draw[thick] (O)  to  node[midway,sloped]{$>$} ++ (225:1.5);

\draw[thick,green] (O)  to  node[midway,sloped]{$>$}  ++ (1.2,0) ;
\draw[->,>=stealth,very thick] (O) ++ (2.1,0) -- ++ (1,0);

\coordinate (B) at (9.5,1.5);

\path (B) ++ (180:1.5) coordinate (P1) ++ (240:1) coordinate (P2) ++ (225:1) coordinate (P3) ++ (210:1) coordinate (P4);

\draw[thick] (P1) --node[midway,sloped]{$>$}++ (130:1.5) ++(60:0.75) node {${j}_{1}$};
\draw[thick] (P2) --node[midway,sloped]{$>$}++ (160:1.5) ++(60:0.75) node {${j}_{2}$};

\draw[thick] (P4) --node[midway,sloped]{$>$}++ (185:1.5) ++(60:0.35) node {${j}_{P-1}$};

\draw[thick,green] (B) --node[midway,sloped]{$>$}node[midway,sloped]{$>$}node[above,black] {$j$} (P1);

\draw[thick] (P4) --node[midway,sloped]{$>$}node[near end,left] {${j}_{P}$}++(220:1.5) ++ (300:0.75);

\path (P3) ++ (100:0.75) ++ (-60:0.2) coordinate (O3);
\path (P4) ++ (60:0.35) ++ (120:0.2) coordinate (O4);
\draw [thick, loosely dotted] (O3) -- (O4);

\draw[thick,green] (P1) --node[midway,sloped]{$>$}node[midway,left,black] {${j}_{\imath_1}$} (P2) --node[midway,sloped]{$>$}node[below,black] {${j}_{\imath_2}$} (P3) --node[midway,sloped]{$>$}node[below,black] {${j}_{\imath_p}$} (P4);

\end{tikzpicture}
\caption{The illustration of recoupling spins for $|\mathcal{I}_P;j,m,k\rangle$.}
\label{fig:AnyValency-vertex2}
\end{figure}
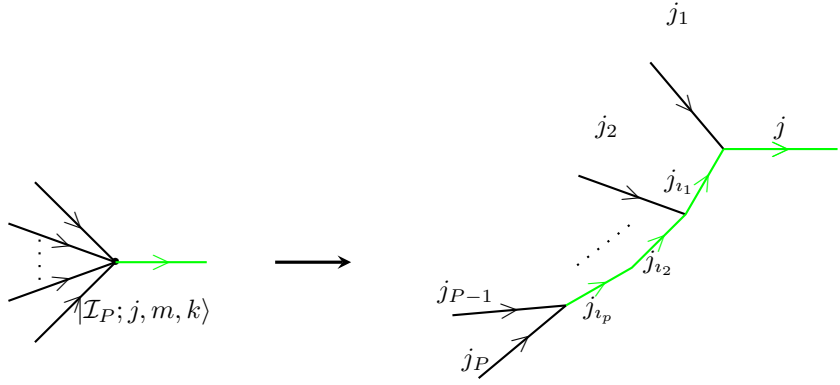
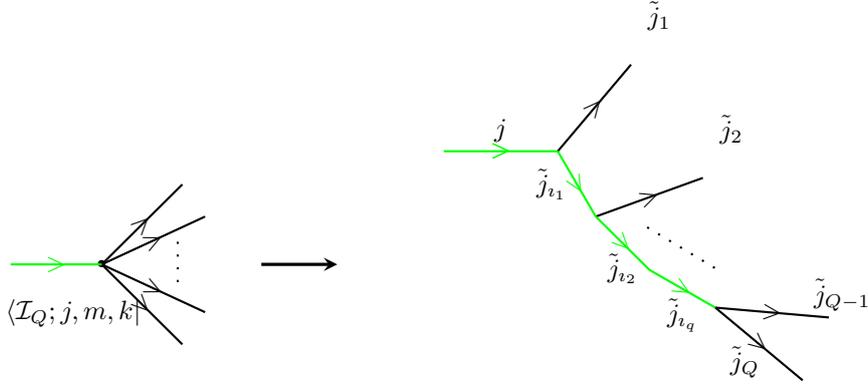
\begin{figure}[htb]
	\centering
	\begin{tikzpicture} [scale=1]

\coordinate  (O) at (0,0);

\node[scale=0.7] at (O) {$\bullet$}  ++ (240:0.75) node [black] {$\langle\mathcal{I}_Q;j,m,k|$};

\draw[thick] (O)  --  node[midway,sloped]{$>$} ++ (45:1.5) ;

\draw[thick] (O)  to  node[midway,sloped]{$>$} ++ (25:1.5) ;

\draw[thick] (O)  --  node[midway,sloped]{$>$} ++ (-25:1.5) ;

\draw[thick] (O)  to  node[midway,sloped]{$>$} ++ (-45:1.5) ;

\coordinate  (O1) at (1,0.3);

\coordinate  (O2) at (1,-0.3);

\draw [thick, loosely dotted] (O1) -- (O2);

\draw[thick,green] (O)  to  node[midway,sloped]{$>$}  ++ (-1.2,0) ;
\draw[->,>=stealth,very thick] (O) ++ (2.1,0) -- ++ (1,0);

\coordinate (B) at (4.5,1.5);

\path (B) ++ (0:1.5) coordinate (P1) ++ (300:1) coordinate (P2) ++ (315:1) coordinate (P3) ++ (330:1) coordinate (P4);

\draw[thick] (P1) --node[midway,sloped]{$>$}++ (50:1.5) ++(60:0.75) node {$\tilde{j}_{1}$};
\draw[thick] (P2) --node[midway,sloped]{$>$}++ (20:1.5) ++(60:0.75) node {$\tilde{j}_{2}$};

\draw[thick] (P4) --node[midway,sloped]{$>$}++ (355:1.5) ++(60:0.35) node {$\tilde{j}_{Q-1}$};

\draw[thick,green] (B) --node[midway,sloped]{$>$}node[midway,sloped]{$>$}node[above,black] {$j$} (P1);

\draw[thick] (P4) --node[midway,sloped]{$>$} node[near end,left=0.15] {$\tilde{j}_{Q}$}++(320:1.5) ++ (300:0.75);

\path (P3) ++ (100:0.75) ++ (-60:0.2) coordinate (O3);
\path (P4) ++ (60:0.35) ++ (120:0.2) coordinate (O4);
\draw [thick, loosely dotted] (O3) -- (O4);

\draw[thick,green] (P1) --node[midway,sloped]{$>$}node[midway,left,black] {$\tilde{j}_{\imath_1}$} (P2) --node[midway,sloped]{$>$}node[below,black] {$\tilde{j}_{\imath_2}$} (P3) --node[midway,sloped]{$>$}node[below,black] {$\tilde{j}_{\imath_q}$} (P4);

\end{tikzpicture}
\caption{The illustration of recoupling spins for $\langle\mathcal{I}_Q;j,m,k|$.}
\label{fig:AnyValency-vertex1}
\end{figure}
One can further insert recoupling edges adapting to a recoupling scheme for $|\mathcal{I}_{P},j,m\rangle$ step by step.
Following the re-coupling scheme illustrated in Fig.\ref{fig:AnyValency-vertex2} and using character formula for delta function
\begin{eqnarray}
g | j m \rangle \otimes \langle j' m' | g^{-1}
&=&\int_{SU(2)} dh \delta(g^{-1} h) h | j m \rangle \otimes \langle j' m' | h^{-1}
\nonumber
\\ \nonumber
&=&
\sum_{j''} (2j''+1)\int_{SU(2)} dh  \text{tr}^{j''}( hg^{-1}) h | j m \rangle \otimes \langle j' m' | h^{-1}
,
\end{eqnarray}
 one can get
\begin{eqnarray}
&&|\mathcal{I}_{P};j,m,k\rangle:=\int_{SU(2)}dg
\left( \bigotimes_{I=1}^{P}g|j_I,\hat{n}_I\rangle \right)
\otimes \langle j,m|g^{-1}|j,k\rangle\\\nonumber
&=&\sum_{j_{\imath_1}}(2j_{\imath_1}+1)\sum_{m_{\imath_1,k_{\imath_1}}=-j_{\imath_1}}^{j_{\imath_1}}\int_{SU(2)}dg
\bigotimes_{I=2}^{P}g|j_I,\hat{n}_I\rangle\cdot\langle j_{\imath_1},m_{\imath_1}|g^{-1} |j_{\imath_1},k_{\imath_1}\rangle\otimes|j_{\imath_1}k_{\imath_1}m_{\imath_1},j_1\hat{n}_1,jmk\rangle\\\nonumber
&=&\sum_{j_{\imath_1}}\sum_{j_{\imath_2}}(2j_{\imath_1}+1)(2j_{\imath_2}+1) \Bigg(\sum_{m_{\imath_1},k_{\imath_1}=-j_{\imath_1}}^{j_{\imath_1}} \sum_{m_{\imath_2},k_{\imath_2}=-j_{\imath_2}}^{j_{\imath_2}} \int_{SU(2)}dg\bigotimes_{I=3}^{P}g|j_I,\hat{n}_I\rangle\cdot\langle j_{\imath_2},m_{\imath_2}|g^{-1}|j_{\imath_2},k_{\imath_2}\rangle\\\nonumber
&& \otimes|j_{\imath_2}k_{\imath_2}m_{\imath_2},j_2\hat{n}_2,j_{\imath_1}m_{\imath_1}k_{\imath_1}\rangle \otimes|j_{\imath_1}k_{\imath_1}m_{\imath_1},j_1\hat{n}_1,jmk\rangle\Bigg)\\\nonumber
&=&\sum_{j_{\imath_1}}\sum_{j_{\imath_2}}...\sum_{j_{\imath_{p}}} (2j_{\imath_1}+1)(2j_{\imath_2}+1)...(2j_{\imath_p}+1)\\\nonumber
&&\cdot\Bigg(\sum_{m_{\imath_1},k_{\imath_1}=-j_{\imath_1}}^{j_{\imath_1}} \sum_{m_{\imath_2},k_{\imath_2}=-j_{\imath_2}}^{j_{\imath_2}} ...\sum_{m_{\imath_p},k_{\imath_p}=-j_{\imath_p}}^{j_{\imath_p}} |j_{P}\hat{n}_{P},j_{P-1}\hat{n}_{P-1},j_{\imath_p}m_{\imath_p}k_{\imath_p}\rangle \\\nonumber
&& \otimes|j_{\imath_p}k_{\imath_p}m_{\imath_p},j_p\hat{n}_p,j_{\imath_{p-1}}m_{\imath_{p-1}}k_{\imath_{p-1}}\rangle\otimes... \otimes|j_{\imath_2}k_{\imath_{2}}m_{\imath_2},j_2\hat{n}_2,j_{\imath_1}m_{\imath_1}k_{\imath_1}\rangle \otimes|j_{\imath_1}k_{\imath_1}m_{\imath_1},j_1\hat{n}_1,jmk\rangle\Bigg)\\\nonumber
&=&\sum_{j_{\imath_1}}\sum_{j_{\imath_2}}...\sum_{j_{\imath_{p}}} (2j+1)^{-1}\\\nonumber
&&\cdot\sum_{m_{\imath_1},k_{\imath_1}=-j_{\imath_1}}^{j_{\imath_1}} \sum_{m_{\imath_2},k_{\imath_2}=-j_{\imath_2}}^{j_{\imath_2}} ...\sum_{m_{\imath_p},k_{\imath_p}=-j_{\imath_p}}^{j_{\imath_p}} c^{j_{P},j_{P-1},j_{\imath_p}}_{\hat{n}_{P},\hat{n}_{P-1},m_{\imath_p}} c^{j_{\imath_p},j_p,j_{\imath_{p-1}}}_{m_{\imath_p},\hat{n}_p,m_{\imath_{p-1}}} ...c^{j_{\imath_2},j_2,j_{\imath_1}}_{m_{\imath_2},\hat{n}_2,m_{\imath_1}} c^{j_{\imath_1},j_1,j}_{m_{\imath_1},\hat{n}_1,m}\\\nonumber
&&\cdot\Bigg( |j_{P},j_{P-1};j_{\imath_p},k_{\imath_p}\rangle \otimes|j_{\imath_p},k_{\imath_p};j_p;j_{\imath_{p-1}},k_{\imath_{p-1}}\rangle\otimes... \otimes|j_{\imath_2},k_{\imath_2};j_2;j_{\imath_1},k_{\imath_1}\rangle \otimes|j_{\imath_1},k_{\imath_1};j_1;j,k\rangle\Bigg)
\end{eqnarray}
with $p:=P-2$, where we  defined
\begin{eqnarray}
c^{j_{\imath_1},j_1,j}_{m_{\imath_1},\hat{n}_1,m}
&:=&\sum_{k'_{\imath_1}=-j_{\imath_1}}^{j_{\imath_1}}\sum_{k'=-j}^j\langle j_{\imath_1},k'_{\imath_1};j_1;j,k'|j_{\imath_1}k'_{\imath_1}m_{\imath_1},j_1\hat{n}_1,jmk'\rangle,
\end{eqnarray}
\begin{eqnarray}
c^{j_{\imath_2},j_2,j_{\imath_1}}_{m_{\imath_2},\hat{n}_2,m_{\imath_1}}
&:=&\sum_{k'_{\imath_2}=-j_{\imath_2}}^{j_{\imath_2}}\sum_{k'_{\imath_1}=-j_{\imath_1}}^{j_{\imath_1}}\langle j_{\imath_2},k'_{\imath_2};j_2;j_{\imath_1},k_{\imath_1}|j_{\imath_2}k'_{\imath_2}m_{\imath_2},j_2\hat{n}_2,j_{\imath_1}m_{\imath_1}k'_{\imath_1}\rangle,
\end{eqnarray}
\begin{equation}
...,
\end{equation}
\begin{eqnarray}
c^{j_{P},j_{P-1},j_{\imath_p}}_{\hat{n}_{P},\hat{n}_{P-1},m_{\imath_p}}
&:=&\sum_{k'_{\imath_p}=-j_{\imath_p}}^{j_{\imath_p}}\langle j_{P},j_{P-1};j_{\imath_p},k'_{\imath_p}|j_{P}\hat{n}_{P},j_{P-1}\hat{n}_{P-1},j_{\imath_p}m_{\imath_p}k'_{\imath_p}\rangle,
\end{eqnarray}
and
\begin{eqnarray}
   |j_{\imath_1}k'_{\imath_1}m_{\imath_1},j_1\hat{n}_1,jmk'\rangle&:=& \int_{SU(2)}dg g|j_1,\hat{n}_1\rangle\langle j_{\imath_1}k'_{\imath_1}|g|j_{\imath_1},m_{\imath_1}\rangle\cdot \langle j,m|g^{-1}|j,k'\rangle,
\end{eqnarray}
\begin{eqnarray}
|j_{\imath_2}k'_{\imath_2}m_{\imath_2},j_2\hat{n}_2,j_{\imath_1}m_{\imath_1}k'_{\imath_1}\rangle&:=& \int_{SU(2)}dg g|j_2,\hat{n}_2\rangle\langle j_{\imath_2},k'_{\imath_2}|g|j_{\imath_2},m_{\imath_2}\rangle \cdot\langle j_{\imath_1}m_{\imath_1}|g^{-1}|j_{\imath_1},k'_{\imath_1}\rangle,
\end{eqnarray}
\begin{equation}
...,
\end{equation}
\begin{eqnarray}
|j_{P}\hat{n}_{P},j_{P-1}\hat{n}_{P-1},j_{\imath_p}m_{\imath_p}k'_{\imath_p}\rangle&:=& \int_{SU(2)}dgg|j_P,\hat{n}_P\rangle\otimes g|j_{P-1},\hat{n}_{P-1}\rangle \langle j_{\imath_p},m_{\imath_p}|g^{-1}|j_{\imath_p},k'_{\imath_p}\rangle,
\end{eqnarray}
with   $|j',j'';j,k\rangle:=\sum_{m',m''}C^{j'j''j}_{m'm''k}|j',m'\rangle \otimes|j'',m''\rangle$,  $|j',k';j'';j,k\rangle:=\sum_{m''}C^{j'j''j}_{k'm''k}|j'',m''\rangle$.
Similarly, as the  illustration in Fig.\ref{fig:AnyValency-vertex1}, one can expand
\begin{eqnarray}
&&\langle\mathcal{I}_{Q};j,m,k|:=\int_{SU(2)}dh\langle j,k|h| j,m\rangle\bigotimes_{J=1}^{Q} \langle \tilde{j}_J,\hat{\tilde{n}}_J|h^{-1}
\\\nonumber
&=&\sum_{\tilde{j}_{\imath_1}}(2\tilde{j}_{\imath_1}+1)\Bigg(\sum_{\tilde{m}_{\imath_1},\tilde{k}_{\imath_1}=-\tilde{j}_{\imath_1}}^{\tilde{j}_{\imath_1}}\int_{SU(2)}dh\langle \tilde{j}_{\imath_1},\tilde{k}_{\imath_1}| h| \tilde{j}_{\imath_1},\tilde{m}_{\imath_1}\rangle\bigotimes_{J=2}^{Q} \langle \tilde{j}_J,\hat{\tilde{n}}_J|h^{-1} \otimes\langle \tilde{j}_{\imath_1}\tilde{m}_{\imath_1}\tilde{k}_{\imath_1},\tilde{j}_1\hat{\tilde{n}}_1,jkm|\Bigg)\\\nonumber
&=&\sum_{\tilde{j}_{\imath_1}}\sum_{\tilde{j}_{\imath_2}}...\sum_{\tilde{j}_{\imath_{q}}} (2j+1)^{-1}\\\nonumber
&&\cdot\sum_{\tilde{m}_{\imath_1},\tilde{k}_{\imath_1}=-\tilde{j}_{\imath_1}}^{\tilde{j}_{\imath_1}} \sum_{\tilde{m}_{\imath_2}, \tilde{k}_{\imath_2}=-\tilde{j}_{\imath_2}}^{\tilde{j}_{\imath_2}} ...\sum_{\tilde{m}_{\imath_q}, \tilde{k}_{\imath_q}=-\tilde{j}_{\imath_q}}^{\tilde{j}_{\imath_q}} c^{\tilde{j}_{Q},\tilde{j}_{Q-1},\tilde{j}_{\imath_q}}_{\hat{\tilde{n}}_{Q},\hat{\tilde{n}}_{Q-1},\tilde{m}_{\imath_q}} c^{\tilde{j}_{\imath_q},\tilde{j}_q,\tilde{j}_{\imath_{q-1}}}_{\tilde{m}_{\imath_q},\hat{\tilde{n}}_q,\tilde{m}_{\imath_{q-1}}} ...c^{\tilde{j}_{\imath_2},\tilde{j}_2,\tilde{j}_{\imath_1}}_{\tilde{m}_{\imath_2},\hat{\tilde{n}}_2,\tilde{m}_{\imath_1}} c^{\tilde{j}_{\imath_1},\tilde{j}_1,j}_{\tilde{m}_{\imath_1},\hat{\tilde{n}}_1,m}\\\nonumber
&&\cdot  \Bigg( \langle\tilde{j}_{Q},\tilde{j}_{Q-1};\tilde{j}_{\imath_q},\tilde{k}_{\imath_q}| \otimes\langle \tilde{j}_{\imath_q},\tilde{k}_{\imath_q};\tilde{j}_q;\tilde{j}_{\imath_{q-1}},\tilde{k}_{\imath_{q-1}}| \otimes... \otimes\langle \tilde{j}_{\imath_2},\tilde{k}_{\imath_2};\tilde{j}_2;\tilde{j}_{\imath_1},\tilde{k}_{\imath_1}| \otimes\langle \tilde{j}_{\imath_1},\tilde{k}_{\imath_1};\tilde{j}_1;j,k|\Bigg)
\end{eqnarray}
with $q=Q-2$, where we defined  
\begin{eqnarray}
c^{\tilde{j}_{\imath_1},\tilde{j}_1,j}_{\tilde{m}_{\imath_1},\hat{\tilde{n}}_1,m}&:=&\sum_{\tilde{k}'_{\imath_1},k'}
\langle \tilde{j}_{\imath_1}\tilde{m}_{\imath_1}\tilde{k}'_{\imath_1},\tilde{j}_1\hat{\tilde{n}}_1,jk'm|\tilde{j}_{\imath_1},\tilde{k}'_{\imath_1};\tilde{j}_1;j,k'\rangle,
\end{eqnarray}
\begin{eqnarray}
c^{\tilde{j}_{\imath_2},\tilde{j}_2,\tilde{j}_{\imath_1}} _{\tilde{m}_{\imath_2},\hat{\tilde{n}}_2,\tilde{m}_{\imath_1}}&:=&\sum_{\tilde{k}'_{\imath_1},\tilde{k}'_{\imath_2}}\langle\tilde{ j}_{\imath_2}\tilde{m}_{\imath_2}\tilde{k}'_{\imath_2},\tilde{j}_2\hat{\tilde{n}}_2,\tilde{j}_{\imath_1}\tilde{k}'_{\imath_1}\tilde{m}_{\imath_1}|\tilde{j}_{\imath_2},\tilde{k}'_{\imath_2};\tilde{j}_2;\tilde{j}_{\imath_1},\tilde{k}'_{\imath_1}\rangle,
\end{eqnarray}
\begin{equation}
...,
\end{equation}
\begin{eqnarray}
c^{\tilde{j}_{Q},\tilde{j}_{Q-1},\tilde{j}_{\imath_q}}_{\hat{\tilde{n}}_{Q},\hat{\tilde{n}}_{Q-1},\tilde{m}_{\imath_q}}&:=&\sum_{\tilde{k}'_{\imath_q}}\langle\tilde{ j}_{Q}\hat{\tilde{n}}_{Q},\tilde{j}_{Q-1}\hat{\tilde{n}}_{Q-1},\tilde{j}_{\imath_q}\tilde{k}'_{\imath_q}\tilde{m}_{\imath_q}|\tilde{j}_{Q},\tilde{j}_{Q-1};\tilde{j}_{\imath_q},\tilde{k}'_{\imath_q}\rangle,
\end{eqnarray}
and
\begin{equation}
\langle \tilde{j}_{\imath_1}\tilde{m}_{\imath_1}\tilde{k}'_{\imath_1},\tilde{j}_1\hat{\tilde{n}}_1,jk'm|:= \int_{SU(2)}dg
\otimes
\langle j,k'|g|j,m\rangle\langle \tilde{j}_{\imath_1},\tilde{m}_{\imath_1}|g^{-1}|\tilde{j}_{\imath_1},\tilde{k}'_{\imath_1}\rangle\otimes\langle \tilde{j}_1,\hat{\tilde{n}}_1|g^{-1},
\end{equation}
\begin{equation}
\langle\tilde{ j}_{\imath_2}\tilde{m}_{\imath_2}\tilde{k}'_{\imath_2},\tilde{j}_2\hat{\tilde{n}}_2,\tilde{j}_{\imath_1}\tilde{k}'_{\imath_1}\tilde{m}_{\imath_1}|:=\int_{SU(2)}dg
\langle \tilde{ j}_{\imath_1},\tilde{k}'_{\imath_1}|
g|\tilde{ j}_{\imath_1},\tilde{m}_{\imath_1}\rangle\langle \tilde{j}_{\imath_2},\tilde{m}_{\imath_2}|g^{-1}|\tilde{j}_{\imath_2},\tilde{k}'_{\imath_2}\rangle\otimes\langle\tilde{ j}_2,\hat{\tilde{n}}_2|g^{-1},
\end{equation}
\begin{equation}
...,
\end{equation}
\begin{equation}
\langle\tilde{ j}_{Q}\hat{\tilde{n}}_{Q},\tilde{j}_{Q-1}\hat{\tilde{n}}_{Q-1},\tilde{j}_{\imath_q}\tilde{k}'_{\imath_q}\tilde{m}_{\imath_q}|:= \int_{SU(2)}dg
\langle\tilde{ j}_{\imath_q},\tilde{k}'_{\imath_q}|
g|\tilde{ j}_{\imath_q},\tilde{m}_{\imath_q}\rangle\langle \tilde{j}_{Q},\hat{\tilde{n}}_{Q}|g^{-1}\otimes\langle\tilde{ j}_{Q-1},\hat{\tilde{n}}_{Q-1}|g^{-1}.
\end{equation}
Then, one has
\begin{eqnarray}
&&|\mathcal{I}\rangle\\\nonumber
   &=&\sum_{j}(2j+1)\Bigg(\sum_{m,k=-j}^j |\mathcal{I}_{P};j,m,k\rangle\langle\mathcal{I}_{Q};j,m,k|\Bigg)\\\nonumber
   &=&\sum_{j}\sum_{j_{\imath_1}}\sum_{j_{\imath_2}}...\sum_{j_{\imath_{p}}} \sum_{\tilde{j}_{\imath_1}}\sum_{\tilde{j}_{\imath_2}}...\sum_{\tilde{j}_{\imath_{q}}} (2j+1)^{-1} \\\nonumber
   &&\cdot c^{\mathcal{I}_{P},\mathcal{I}_{Q}}_{j,j_\imath,\tilde{j}_\imath}\cdot\sum_{k=-j}^j\sum_{\tilde{k}_{\imath_1}=-\tilde{j}_{\imath_1}}^{\tilde{j}_{\imath_1}} \sum_{\tilde{k}_{\imath_2}=-\tilde{j}_{\imath_2}}^{\tilde{j}_{\imath_2}} ...\sum_{\tilde{k}_{\imath_q}=-\tilde{j}_{\imath_q}}^{\tilde{j}_{\imath_q}}  \sum_{k_{\imath_1}=-j_{\imath_1}}^{j_{\imath_1}} \sum_{k_{\imath_2}=-j_{\imath_2}}^{j_{\imath_2}} ...\sum_{k_{\imath_p}=-j_{\imath_p}}^{j_{\imath_p}}\Bigg(  |j_{P},j_{P-1};j_{\imath_p},k_{\imath_p}\rangle  \\\nonumber
   &&\otimes|j_{\imath_p},k_{\imath_p};j_p;j_{\imath_{p-1}},k_{\imath_{p-1}}\rangle\otimes...  \otimes|j_{\imath_2},k_{\imath_2};j_2;j_{\imath_1},k_{\imath_1}\rangle \otimes|j_{\imath_1},k_{\imath_1};j_1;j,k\rangle\langle\tilde{j}_{Q},\tilde{j}_{Q-1};\tilde{j}_{\imath_q},\tilde{k}_{\imath_q}| \\\nonumber
   &&\otimes\langle \tilde{j}_{\imath_q},\tilde{k}_{\imath_q};\tilde{j}_q;\tilde{j}_{\imath_{q-1}},\tilde{k}_{\imath_{q-1}}| \otimes... \otimes\langle \tilde{j}_{\imath_2},\tilde{k}_{\imath_2};\tilde{j}_2;\tilde{j}_{\imath_1},\tilde{k}_{\imath_1}| \otimes\langle \tilde{j}_{\imath_1},\tilde{k}_{\imath_1};\tilde{j}_1;j,k|\Bigg),
\end{eqnarray}
where
\begin{eqnarray}
 && c^{\mathcal{I}_{P},\mathcal{I}_{Q}}_{j,j_\imath,\tilde{j}_\imath}\\\nonumber
 &:=&\sum_{m=-j}^j\sum_{\tilde{m}_{\imath_1}=-\tilde{j}_{\imath_1}}^{\tilde{j}_{\imath_1}} \sum_{\tilde{m}_{\imath_2}=-\tilde{j}_{\imath_2}}^{\tilde{j}_{\imath_2}} ...\sum_{\tilde{m}_{\imath_q}=-\tilde{j}_{\imath_q}}^{\tilde{j}_{\imath_q}}  \sum_{m_{\imath_1}=-j_{\imath_1}}^{j_{\imath_1}} \sum_{m_{\imath_2}=-j_{\imath_2}}^{j_{\imath_2}} ...\sum_{m_{\imath_p}=-j_{\imath_p}}^{j_{\imath_p}} c^{\tilde{j}_{Q},\tilde{j}_{Q-1},\tilde{j}_{\imath_q}}_{\hat{\tilde{n}}_{Q},\hat{\tilde{n}}_{Q-1},\tilde{m}_{\imath_q}} \\\nonumber
&&\cdot  c^{\tilde{j}_{\imath_q},\tilde{j}_q,\tilde{j}_{\imath_{q-1}}}_{\tilde{m}_{\imath_q},\hat{\tilde{n}}_q,\tilde{m}_{\imath_{q-1}}}  ...c^{\tilde{j}_{\imath_2},\tilde{j}_2,\tilde{j}_{\imath_1}}_{\tilde{m}_{\imath_2},\hat{\tilde{n}}_2,\tilde{m}_{\imath_1}} c^{\tilde{j}_{\imath_1},\tilde{j}_1,j}_{\tilde{m}_{\imath_1},\hat{\tilde{n}}_1,m} c^{j_{P},j_{P-1},j_{\imath_p}}_{\hat{n}_{P},\hat{n}_{P-1},m_{\imath_p}} c^{j_{\imath_p},j_p,j_{\imath_{p-1}}}_{m_{\imath_p},\hat{n}_p,m_{\imath_{p-1}}} ...c^{j_{\imath_2},j_2,j_{\imath_1}}_{m_{\imath_2},\hat{n}_2,m_{\imath_1}} c^{j_{\imath_1},j_1,j}_{m_{\imath_1},\hat{n}_1,m}.
\end{eqnarray}

\end{appendix}

\bibliographystyle{bib-style}
\bibliography{ref}

\providecommand{\href}[2]{#2}\begingroup\raggedright\begin{thebibliography}{10}

\bibitem{30years}
A.~Ashtekar and J.~Pulliny, {\em Loop quantum gravity: The first 30 years}.
\newblock World Scientific Publishing Co. Pte Ltd, Singapore, Mar., 2017.

\bibitem{Ashtekar:2004eh}
A.~Ashtekar and J.~Lewandowski, ``{Background independent quantum gravity: A
  Status report},'' Class. Quant. Grav. {\bf 21} (2004) R53,
  \href{http://arXiv.org/abs/gr-qc/0404018}{{\texttt{arXiv:gr-qc/0404018}}}.

\bibitem{thiemann2008modern}
T.~Thiemann, {\em Modern Canonical Quantum General Relativity}.
\newblock Cambridge University Press, 2007.

\bibitem{rovelli_vidotto_2014}
C.~Rovelli and F.~Vidotto, {\em Covariant Loop Quantum Gravity: An Elementary
  Introduction to Quantum Gravity and Spinfoam Theory}.
\newblock Cambridge University Press, 2014.

\bibitem{rovelli2007quantum}
C.~Rovelli, {\em Quantum gravity}.
\newblock Cambridge university press, 2007.

\bibitem{Han2005FUNDAMENTAL}
M.~Han, M.~A. Yongge, and W.~Huang, ``FUNDAMENTAL STRUCTURE OF LOOP QUANTUM
  GRAVITY,'' International Journal of Modern Physics D {\bf 16} (2005), no.~09,
  1397--1474.

\bibitem{Rovelli:2010km}
C.~Rovelli and S.~Speziale, ``{On the geometry of loop quantum gravity on a
  graph},'' Phys. Rev. {\bf D82} (2010) 044018,
\href{http://arXiv.org/abs/1005.2927}{{\texttt{arXiv:1005.2927}}}.

\bibitem{Freidel:2010bw}
L.~Freidel and S.~Speziale, ``{From twistors to twisted geometries},'' Phys.
  Rev. D {\bf 82} (2010) 084041,
  \href{http://arXiv.org/abs/1006.0199}{{\texttt{arXiv:1006.0199}}}.

\bibitem{PhysRevD.82.084040}
L.~Freidel and S.~Speziale, ``Twisted geometries: A geometric parametrization
  of SU(2) phase space,'' Phys. Rev. D {\bf 82} (Oct, 2010) 084040.

\bibitem{PhysRevD.103.086016}
G.~Long and C.-Y. Lin, ``Geometric parametrization of $\textsc{SO(D+1)}$ phase
  space of all dimensional loop quantum gravity,'' Phys. Rev. D {\bf 103} (Apr,
  2021) 086016.

\bibitem{Long:2023ivt}
G.~Long, ``{Parametrization of holonomy-flux phase space in the Hamiltonian
  formulation of $SO(N)$ gauge field theory with $SO(D+1)$ loop quantum gravity
  as an exemplification},''
  \href{http://arXiv.org/abs/2307.05542}{{\texttt{arXiv:2307.05542}}}.

\bibitem{regge}
T.~Regge, ``{General relativity without coordinates},'' Nuovo Cim. {\bf 19}
  (1961)
558--571.

\bibitem{Alesci:2015wla}
E.~Alesci, M.~Assanioussi, J.~Lewandowski, and I.~Makinen, ``{Hamiltonian
  operator for loop quantum gravity coupled to a scalar field},'' Phys. Rev.
  {\bf D91} (2015), no.~12, 124067,
\href{http://arXiv.org/abs/1504.02068}{{\texttt{arXiv:1504.02068}}}.

\bibitem{Assanioussi:2015gka}
M.~Assanioussi, J.~Lewandowski, and I.~Makinen, ``{New scalar constraint
  operator for loop quantum gravity},'' Phys. Rev. {\bf D92} (2015), no.~4,
  044042,
\href{http://arXiv.org/abs/1506.00299}{{\texttt{arXiv:1506.00299}}}.

\bibitem{Zhang:2021qul}
C.~Zhang, S.~Song, and M.~Han, ``{First-Order Quantum Correction in Coherent
  State Expectation Value of Loop-Quantum-Gravity Hamiltonian},'' Phys. Rev. D
  {\bf 105} (2022) 064008,
  \href{http://arXiv.org/abs/2102.03591}{{\texttt{arXiv:2102.03591}}}.

\bibitem{Yang:2015zda}
J.~Yang and Y.~Ma, ``{New Hamiltonian constraint operator for loop quantum
  gravity},'' Phys. Lett. B {\bf 751} (2015) 343--347,
  \href{http://arXiv.org/abs/1507.00986}{{\texttt{arXiv:1507.00986}}}.

\bibitem{Perez:2012wv}
A.~Perez, ``{The Spin Foam Approach to Quantum Gravity},'' Living Rev. Rel.
  {\bf 16} (2013) 3,
  \href{http://arXiv.org/abs/1205.2019}{{\texttt{arXiv:1205.2019}}}.

\bibitem{Han:2021kll}
M.~Han, Z.~Huang, H.~Liu, and D.~Qu, ``{Complex critical points and curved
  geometries in four-dimensional Lorentzian spinfoam quantum gravity},'' Phys.
  Rev. D {\bf 106} (2022), no.~4, 044005,
  \href{http://arXiv.org/abs/2110.10670}{{\texttt{arXiv:2110.10670}}}.

\bibitem{Han:2019vpw}
M.~Han and H.~Liu, ``{Effective Dynamics from Coherent State Path Integral of
  Full Loop Quantum Gravity},'' Phys. Rev. {\bf D101} (2020), no.~4, 046003,
\href{http://arXiv.org/abs/1910.03763}{{\texttt{arXiv:1910.03763}}}.

\bibitem{Han:2020chr}
M.~Han and H.~Liu, ``{Semiclassical limit of new path integral formulation from
  reduced phase space loop quantum gravity},'' Phys. Rev. D {\bf 102} (2020),
  no.~2, 024083,
  \href{http://arXiv.org/abs/2005.00988}{{\texttt{arXiv:2005.00988}}}.

\bibitem{Long:2021izw}
G.~Long and Y.~Ma, ``{Effective dynamics of weak coupling loop quantum
  gravity},'' Phys. Rev. D {\bf 105} (2022), no.~4, 044043,
  \href{http://arXiv.org/abs/2111.11844}{{\texttt{arXiv:2111.11844}}}.

\bibitem{Ashtekar:1996eg}
A.~Ashtekar and J.~Lewandowski, ``{Quantum theory of geometry. 1: Area
  operators},'' Class. Quant. Grav. {\bf 14} (1997) A55--A82,
  \href{http://arXiv.org/abs/gr-qc/9602046}{{\texttt{arXiv:gr-qc/9602046}}}.

\bibitem{Ashtekar:1997fb}
A.~Ashtekar and J.~Lewandowski, ``{Quantum theory of geometry. 2. Volume
  operators},'' Adv. Theor. Math. Phys. {\bf 1} (1998) 388--429,
  \href{http://arXiv.org/abs/gr-qc/9711031}{{\texttt{arXiv:gr-qc/9711031}}}.

\bibitem{Ma:2010fy}
Y.~Ma, C.~Soo, and J.~Yang, ``{New length operator for loop quantum gravity},''
  Phys. Rev. D {\bf 81} (2010) 124026,
  \href{http://arXiv.org/abs/1004.1063}{{\texttt{arXiv:1004.1063}}}.

\bibitem{Bianchi:2008es}
E.~Bianchi, ``{The Length operator in Loop Quantum Gravity},'' Nucl. Phys. B
  {\bf 807} (2009) 591--624,
  \href{http://arXiv.org/abs/0806.4710}{{\texttt{arXiv:0806.4710}}}.

\bibitem{long2020operators}
G.~Long and Y.~Ma, ``{General geometric operators in all dimensional loop
  quantum gravity},'' Phys. Rev. D {\bf 101} (2020), no.~8, 084032,
  \href{http://arXiv.org/abs/2003.03952}{{\texttt{arXiv:2003.03952}}}.

\bibitem{Thomas2001Gauge}
T.~Thiemann, ``Gauge field theory coherent states (GCS): I. General
  properties,'' Classical and Quantum Gravity {\bf 18} (2001), no.~11,.

\bibitem{2001Gauge}
T.~Thiemann and O.~Winkler, ``Gauge field theory coherent states (GCS): II.
  Peakedness properties,'' Classical and Quantum Gravity {\bf 18} (2001),
  no.~14, 2561--2636.

\bibitem{Bianchi:2009ky}
E.~Bianchi, E.~Magliaro, and C.~Perini, ``{Coherent spin-networks},'' Phys.
  Rev. D {\bf 82} (2010) 024012,
  \href{http://arXiv.org/abs/0912.4054}{{\texttt{arXiv:0912.4054}}}.

\bibitem{Calcinari_2020}
A.~Calcinari, L.~Freidel, E.~Livine, and S.~Speziale, ``Twisted geometries
  coherent states for loop quantum gravity,'' Classical and Quantum Gravity
  {\bf 38} (Dec, 2020) 025004.

\bibitem{Livine:2006xk}
E.~R. Livine and D.~R. Terno, ``{Reconstructing quantum geometry from quantum
  information: Area renormalisation, coarse-graining and entanglement on spin
  networks},''
\href{http://arXiv.org/abs/gr-qc/0603008}{{\texttt{arXiv:gr-qc/0603008}}}.

\bibitem{Donnelly:2008vx}
W.~Donnelly, ``{Entanglement entropy in loop quantum gravity},'' Phys. Rev. D
  {\bf 77} (2008) 104006,
  \href{http://arXiv.org/abs/0802.0880}{{\texttt{arXiv:0802.0880}}}.

\bibitem{Donnelly:2011hn}
W.~Donnelly, ``{Decomposition of entanglement entropy in lattice gauge
  theory},'' Phys. Rev. {\bf D85} (2012) 085004,
\href{http://arXiv.org/abs/1109.0036}{{\texttt{arXiv:1109.0036}}}.

\bibitem{Donnelly:2014gva}
W.~Donnelly, ``{Entanglement entropy and nonabelian gauge symmetry},'' Class.
  Quant. Grav. {\bf 31} (2014), no.~21, 214003,
\href{http://arXiv.org/abs/1406.7304}{{\texttt{arXiv:1406.7304}}}.

\bibitem{Bianchi:2015fra}
E.~Bianchi, L.~Hackl, and N.~Yokomizo, ``{Entanglement entropy of squeezed
  vacua on a lattice},'' Phys. Rev. {\bf D92} (2015), no.~8, 085045,
\href{http://arXiv.org/abs/1507.01567}{{\texttt{arXiv:1507.01567}}}.

\bibitem{Delcamp:2016eya}
C.~Delcamp, B.~Dittrich, and A.~Riello, ``{On entanglement entropy in
  non-Abelian lattice gauge theory and 3D quantum gravity},'' JHEP {\bf 11}
  (2016) 102,
\href{http://arXiv.org/abs/1609.04806}{{\texttt{arXiv:1609.04806}}}.

\bibitem{Livine:2017fgq}
E.~R. Livine, ``{Intertwiner Entanglement on Spin Networks},'' Phys. Rev. {\bf
  D97} (2018), no.~2, 026009,
\href{http://arXiv.org/abs/1709.08511}{{\texttt{arXiv:1709.08511}}}.

\bibitem{Baytas:2018wjd}
B.~Baytas, E.~Bianchi, and N.~Yokomizo, ``{Gluing polyhedra with entanglement
  in loop quantum gravity},'' Phys. Rev. {\bf D98} (2018), no.~2, 026001,
\href{http://arXiv.org/abs/1805.05856}{{\texttt{arXiv:1805.05856}}}.

\bibitem{Chen:2022rty}
Q.~Chen and E.~R. Livine, ``{Intertwiner entanglement excitation and holonomy
  operator},'' Class. Quant. Grav. {\bf 39} (2022), no.~21, 215013,
  \href{http://arXiv.org/abs/2204.03093}{{\texttt{arXiv:2204.03093}}}.

\bibitem{PhysRevD.104.046014}
G.~Long, C.~Zhang, and X.~Zhang, ``Superposition type coherent states in all
  dimensional loop quantum gravity,'' Phys. Rev. D {\bf 104} (Aug, 2021)
  046014.

\bibitem{1994The}
B.~Hall, ``The Segal-Bargmann "Coherent State" Transform for Compact Lie
  Groups,'' Journal of Functional Analysis {\bf 122} (1994), no.~1, 103--151.

\bibitem{Long:2021lmd}
G.~Long, X.~Zhang, and C.~Zhang, ``{Twisted geometry coherent states in all
  dimensional loop quantum gravity: I. Construction and Peakedness
  properties},''
  \href{http://arXiv.org/abs/2110.01317}{{\texttt{arXiv:2110.01317}}}.

\bibitem{Long:2022cex}
G.~Long, ``{Twisted geometry coherent states in all dimensional loop quantum
  gravity. II. Ehrenfest property},'' Phys. Rev. D {\bf 106} (2022), no.~6,
  066021, \href{http://arXiv.org/abs/2204.03056}{{\texttt{arXiv:2204.03056}}}.

\bibitem{Livine:2007vk}
E.~R. Livine and S.~Speziale, ``{A New spinfoam vertex for quantum gravity},''
  Phys. Rev. D {\bf 76} (2007) 084028,
  \href{http://arXiv.org/abs/0705.0674}{{\texttt{arXiv:0705.0674}}}.

\bibitem{long2019coherent}
G.~Long, C.-Y. Lin, and Y.~Ma, ``Coherent intertwiner solution of simplicity
  constraint in all dimensional loop quantum gravity,'' Physical Review D {\bf
  100} (2019), no.~6, 064065.

\bibitem{Perelomov:1986tf}
A.~M. Perelomov, {\em {Generalized coherent states and their applications}}.
\newblock 1986.

\bibitem{Long:2020euh}
G.~Long and N.~Bodendorfer, ``{Perelomov-type coherent states of SO($D+1$) in
  all-dimensional loop quantum gravity},'' Phys. Rev. D {\bf 102} (2020),
  no.~12, 126004,
  \href{http://arXiv.org/abs/2006.13122}{{\texttt{arXiv:2006.13122}}}.

\bibitem{PhysRevD.83.044035}
E.~Bianchi, P.~Don\'a, and S.~Speziale, ``Polyhedra in loop quantum gravity,''
  Phys. Rev. D {\bf 83} (Feb, 2011) 044035.

\bibitem{Long:2020agv}
G.~Long and Y.~Ma, ``{Polytopes in all dimensional loop quantum gravity},''
  Eur. Phys. J. C {\bf 82} (2022), no.~41,.

\bibitem{Freidel:2002xb}
L.~Freidel and E.~R. Livine, ``{Spin networks for noncompact groups},'' J.
  Math. Phys. {\bf 44} (2003) 1322--1356,
  \href{http://arXiv.org/abs/hep-th/0205268}{{\texttt{arXiv:hep-th/0205268}}}.

\bibitem{Charles:2016xwc}
C.~Charles and E.~R. Livine, ``{The Fock Space of Loopy Spin Networks for
  Quantum Gravity},'' Gen. Rel. Grav. {\bf 48} (2016), no.~8, 113,
  \href{http://arXiv.org/abs/1603.01117}{{\texttt{arXiv:1603.01117}}}.

\bibitem{Deser:1983tn}
S.~Deser, R.~Jackiw, and G.~'t~Hooft, ``{Three-Dimensional Einstein Gravity:
  Dynamics of Flat Space},'' Annals Phys. {\bf 152} (1984) 220.

\bibitem{Freidel:2005me}
L.~Freidel and E.~R. Livine, ``{3D Quantum Gravity and Effective Noncommutative
  Quantum Field Theory},'' Phys. Rev. Lett. {\bf 96} (2006) 221301,
  \href{http://arXiv.org/abs/hep-th/0512113}{{\texttt{arXiv:hep-th/0512113}}}.

\bibitem{Livine:2008iq}
E.~R. Livine and D.~R. Terno, ``{The Entropic boundary law in BF theory},''
  Nucl. Phys. B {\bf 806} (2009) 715--734,
  \href{http://arXiv.org/abs/0805.2536}{{\texttt{arXiv:0805.2536}}}.

\bibitem{Ashtekar:1997yu}
A.~Ashtekar, J.~Baez, A.~Corichi, and K.~Krasnov, ``{Quantum geometry and black
  hole entropy},'' Phys. Rev. Lett. {\bf 80} (1998) 904--907,
  \href{http://arXiv.org/abs/gr-qc/9710007}{{\texttt{arXiv:gr-qc/9710007}}}.

\bibitem{Ashtekar:2000eq}
A.~Ashtekar, J.~C. Baez, and K.~Krasnov, ``{Quantum geometry of isolated
  horizons and black hole entropy},'' Adv. Theor. Math. Phys. {\bf 4} (2000)
  1--94,
  \href{http://arXiv.org/abs/gr-qc/0005126}{{\texttt{arXiv:gr-qc/0005126}}}.

\bibitem{Engle:2009vc}
J.~Engle, A.~Perez, and K.~Noui, ``{Black hole entropy and SU(2) Chern-Simons
  theory},'' Phys. Rev. Lett. {\bf 105} (2010) 031302,
  \href{http://arXiv.org/abs/0905.3168}{{\texttt{arXiv:0905.3168}}}.

\bibitem{Song:2020arr}
S.~Song, H.~Li, Y.~Ma, and C.~Zhang, ``{Entropy of black holes with arbitrary
  shapes in loop quantum gravity},'' Sci. China Phys. Mech. Astron. {\bf 64}
  (2021), no.~12, 120411,
  \href{http://arXiv.org/abs/2002.08869}{{\texttt{arXiv:2002.08869}}}.

\bibitem{Ghosh:2011fc}
A.~Ghosh and A.~Perez, ``{Black hole entropy and isolated horizons
  thermodynamics},'' Phys. Rev. Lett. {\bf 107} (2011) 241301,
  \href{http://arXiv.org/abs/1107.1320}{{\texttt{arXiv:1107.1320}}}. [Erratum:
  Phys.Rev.Lett. 108, 169901 (2012)].

\bibitem{Song:2022zit}
S.~Song, G.~Long, C.~Zhang, and X.~Zhang, ``{Thermodynamics of isolated
  horizons in loop quantum gravity},'' Phys. Rev. D {\bf 106} (2022), no.~12,
  126007, \href{http://arXiv.org/abs/2205.09984}{{\texttt{arXiv:2205.09984}}}.

\bibitem{Perez:2014ura}
A.~Perez, ``{Statistical and entanglement entropy for black holes in quantum
  geometry},'' Phys. Rev. D {\bf 90} (2014), no.~8, 084015,
  \href{http://arXiv.org/abs/1405.7287}{{\texttt{arXiv:1405.7287}}}. [Addendum:
  Phys.Rev.D 90, 089907 (2014)].

\end{thebibliography}\endgroup

\end{document}